\documentclass[floatfix, superscriptaddress, twocolumn, showpacs, aps, pra, 10pt,amsmath,amssymb]{revtex4-2}
\usepackage{graphicx}% Include figure files
\usepackage{dcolumn}% Align table columns on decimal point
\usepackage{bm}% bold math
\usepackage{stackrel}% http://ctan.org/pkg/stackrel
\usepackage{amsmath}
\usepackage{amsfonts}
\usepackage{amssymb}
\usepackage{amsthm}
\usepackage{epstopdf}
\usepackage{longtable}
\usepackage{array}
\usepackage{pifont}
\usepackage{latexsym}
\usepackage[latin1]{inputenc}
\usepackage{bm}
\usepackage{textcomp}
\usepackage{subfigure}
\usepackage{braket}
\usepackage{ulem}
\usepackage{color, soul}
\usepackage{pifont}
\usepackage{multirow}

\begin{document}
\title{Engineering quantum control with twisted-light fields induced optical transitions}

\author{T. Zanon-Willette}
\affiliation{Sorbonne Université, Observatoire de Paris, Université PSL, CNRS, LERMA, F-75005, Paris, France}
\email{thomas.zanon@sorbonne-universite.fr}
\author{F. Impens}
\affiliation{Instituto de F\'isica, Universidade Federal do Rio de Janeiro, Rio de Janeiro, RJ 21941-972, Brazil}
\author{E. Arimondo}
\affiliation{Dipartimento di Fisica E. Fermi, Universit\`a of Pisa -- Lgo. B. Pontecorvo 3, 56127 Pisa, Italy}
\affiliation{INO-CNR, Via G. Moruzzi 1, 56124 Pisa, Italy}
\author{D. Wilkowski}
\address{MajuLab, International Research Laboratory IRL 3654, Université Côte d'Azur, Sorbonne Université, National University of Singapore, Nanyang
Technological University, Singapore,}
\address{Centre for Quantum Technologies, National University of Singapore, 117543 Singapore, Singapore}
\address{School of Physical and Mathematical Sciences, Nanyang Technological University, 637371 Singapore, Singapore}
\author{A.V. Taichenachev}
\address{Novosibirsk State University, ul. Pirogova 2, 630090 Novosibirsk, Russia}
\address{Institute of Laser Physics, Siberian Branch, Russian Academy of Sciences, prosp. Akad. Lavrent'eva 15B, 630090 Novosibirsk, Russia}
\author{V.I. Yudin}
\address{Novosibirsk State University, ul. Pirogova 2, 630090 Novosibirsk, Russia}
\address{Institute of Laser Physics, Siberian Branch, Russian Academy of Sciences, prosp. Akad. Lavrent'eva 15B, 630090 Novosibirsk, Russia}
\address{Novosibirsk State Technical University, prosp. Karla Marksa 20, 630073 Novosibirsk, Russia}

\begin{abstract}
A novel form of quantum control is proposed by applying twisted-light also known as optical vortex beams to drive ultra-narrow atomic transitions in neutral Ca, Mg, Yb, Sr, Hg and Cd bosonic isotopes.
This innovative all-optical spectroscopic method introduces spatially tailored electric and magnetic fields to fully rewrite atomic selection rules reducing simultaneously probe-induced frequency-shifts and additional action of external ac and dc field distortions. A twisted-light focused probe beam produces strong longitudinal electric and magnetic fields along the laser propagation axis which opens the $^{1}$S$_{0}\rightarrow^{3}$P$_{0}$ doubly forbidden clock transition with a high E1M1 two-photon excitation rate. This long-lived clock transition is thus immune to nonscalar electromagnetic perturbations. Zeeman components of the M2 magnetic quadrupole $^{1}$S$_{0}\rightarrow^{3}$P$_{2}$ transition considered for quantum computation and simulation are now selectively driven by transverse or longitudinal field gradients with vanishing electric fields. These field gradients are manipulated by the mutual action of orbital and spin angular momentum of the light beam and are used in presence of tunable vector and tensor polarizabilities. A combination of these two different twisted-light induced clock transitions within a single quantum system, at the same magic wavelength and in presence of a common thermal environment significantly reduces systematic uncertainties. Furthermore, it generates an optical synthetic frequency which efficiently limits the blackbody radiation shift and its variations at room temperature.
Engineering light-matter interaction by optical vortices merged with composite pulses will ultimately benefit to experimental atomic and molecular platforms targeting an optimal coherent control of quantum states, reliant quantum simulation, novel approach to atomic interferometry and precision tests of fundamental theories in physics and high-accuracy optical metrology.
\end{abstract}

\date{\today}

\preprint{APS/123-QED}

\maketitle

\section{Introduction}

\indent The quantum interactions of an atomic system with laser radiation are governed by selection rules and interaction strengths. The selection rules impose specific spatial orientations of electric and magnetic fields driving the optical transition depending on quantum numbers and parity of the initial and final states. Usually, the strength determining the light-matter interaction process is adapted to a specific operation. For example, optical clocks require atom-laser interactions with widely different strengths. At one side, large optical power is required for laser cooling and spatial confinement in optical lattices or tweezer arrays~\cite{Ye:2008,Derevianko:2011,Ludlow:2015,Barredo:2016}. On the opposite side, ultra-narrow dipole-forbidden transitions with very weak strengths are required to increase the clock accuracy for quantum metrology. Electric-quadrupole or octupole transitions have been proposed as potential highly accurate clocks searching for ultra-light dark matter and for testing fundamental theories beyond the standard model~\cite{Flaumbaum:2018,Safronova:2018,Ishiyama:2023,Tang:2023}. At present, the $^{1}$S$_{0}\rightarrow^{3}$P$_{0}$ transition in two-electron atoms has attracted the main attention as a robust optical lattice clock due to the doubly forbidden nature of angular and spin momentum selection rules. The transition is opened by an excited state mixing produced in fermions through the hyperfine interaction~\cite{Boyd:2007} and in bosons by a small dc magnetic field~\cite{Taichenachev:2006,Barber:2006,Baillard:2007,Kulosa:2015}. Unfortunately, this magnetic field leads to complex and hardly controllable frequency shifts, an issue which has limited
the performances of the bosonic clocks so far~\cite{Baillard:2007}. Within a different context, long-lived qubits based on the weakly allowed magnetic quadrupole transition $^{1}$S$_{0}\rightarrow^{3}$P$_{2}$ in two-electron systems are today investigated to form a rich experimental platform for quantum computation and simulation~\cite{Daley:2008,Kato:2012,Onishchenko:2019}. Their advantages are the natural lifetime of hundreds of seconds and the high flexibility in tuning the vector and tensor parts of the atomic polarizability~\cite{Tang:2023,Trautmann:2023,Ishiyama:2023}. Whatever the target is, laser spectroscopy always relies on the standard approach of atoms interacting with an optical plane wave and additional fields if necessary, a very good approximation for the central spot of a laser Gaussian beam but always suffering from probe-induced frequency shifts at various levels of magnitude~\cite{Taichenachev:2006,Santra:2005,Beloy:2021}.
\begin{figure}[t!!]
\center
\resizebox{8.5cm}{!}{\includegraphics[angle=0]{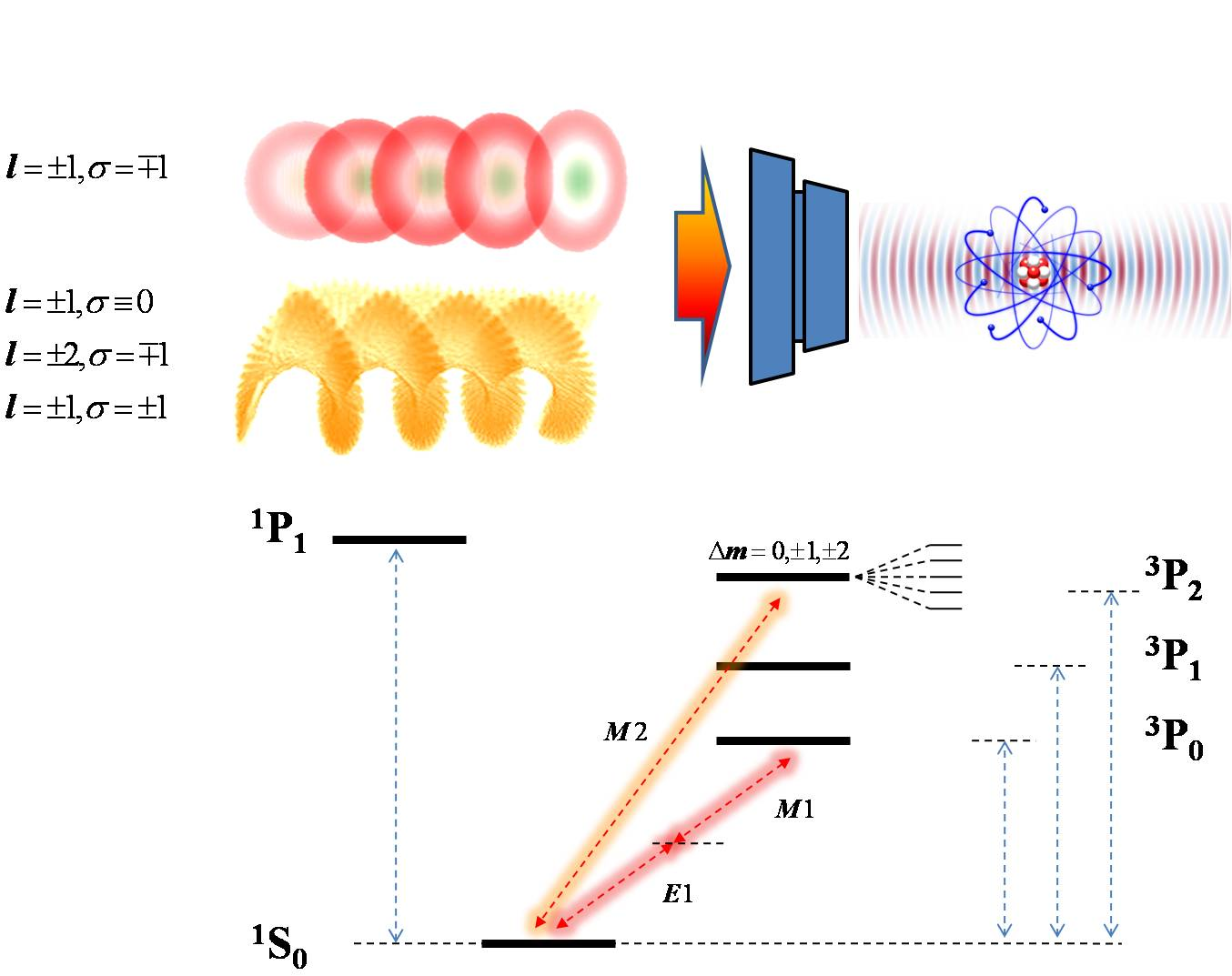}}
\caption{(color online). A two-photon E1M1 clock transition and a magnetic quadrupole M2 transition opened by highly-focused TL fields with monochromatic or bichromatic optical wavelengths. Orbital angular momentum (OAM) integer charge $l$ and spin angular momentum (SAM) helicity $\sigma$ from the structured light can be combined to selectively excite Zeeman $\Delta\textup{m}$ sub-levels with a specific spatial distribution of fields and gradients (top-left $\{l,\sigma\}$ mixture). A linearly polarized TL is introduced by $l=\pm1,\sigma\equiv0$ and circularly polarized TL by $l\neq0,\sigma=\pm1$.}
\label{fig:fig1}
\end{figure}

\indent In this letter, we introduce an innovative all-optical strategy based on an unperturbed bosonic atomic system, where the selection rules are satisfied by an ad-hoc geometry of the electromagnetic fields. Our simple and reliable scheme relies on the use of a single or a superposition of highly-focused twisted-light (TL) running beams to open single to triplet state transitions of bosonic isotopes as depicted in Fig.~\ref{fig:fig1}.

Several benefits are associated to our novel quantum control scheme. The $^{1}$S$_{0}\rightarrow^{3}$P$_{0}$, two-photon optical E1M1 process is activated by using longitudinal electric and magnetic fields generated by a single focused TL beam, a situation not allowed by a single running plane-wave without an external dc magnetic field for state mixing~\cite{Taichenachev:2006,Barber:2006} or requiring a sophisticated two-photon polarization scheme with counter-propagating plane-waves~\cite{Alden:2014}. The use of specific light wave-structures for clock interrogation has been limited so far to very specific cases, such as electric quadrupole (E2) and octupole (E3) transitions~\cite{Schulz:2020,Lange:2022,Verde:2023} leading to a substantial reduction of light-shift in comparison to excitation by a plane-wave. Here, we put forward the benefits of using a light with a tailored-spatial structure to realize a specific multipole clock interrogation allowing control of the amplitude and gradient of the optical field associated to the mixture of polarization and topological charge. We consider specifically TL beams~\cite{Monteiro:2009,Klimov:2012,Quinteiro:2017,Quinteiro:2023}, which have proven a valuable tool with applications ranging from the angular orbital momentum transfer to a single trapped ion~\cite{Schmiegelow:2012,Schmiegelow:2016,Quinteiro:2017-bis} to the engineering of artificial gauge fields~\cite{Shen:2019,Babiker:2019}. TL beams open new excitation channels through longitudinal and transverse components of the electromagnetic fields.
Owing to the large magnetic field gradient associated to the TL beam spatial inhomogeneity, an optical clock with a very high excitation rate is realized for the $^{1}$S$_{0}\rightarrow^{3}$P$_{2}$ magnetic quadrupole transition with vanishing light-shift in the beam center. Compensation of probe-induced shifts can be realized by merging TL beams with composite pulse protocols~\cite{Zanon-Willette:2018}.
Thus, two different TL-induced clock transitions in atoms trapped at a common magic wavelength in an identical thermal environment allow the realization of a synthetic frequency at room temperature that can suppress by several orders of magnitude the blackbody-radiation shift correction, one of the main contribution to the optical clock long term instability~\cite{Yudin:2011,Yudin:2021}.
TL beams are today a mature technique and can thus easily be integrated to the recent generation of optical clocks based on optical tweezers~\cite{Young:2020,Kaufman:2021,Tian:2023}.

\indent The paper contains a main section subdivided into two parts. Paragraph II.A introduces longitudinal electric and magnetic fields produced by a highly focused TL beam which open a two-photon E1M1 optical clock transition. Excitation rates of a monochromatic and a bichromatic two-photon excitation are evaluated with corresponding light-shifts. Paragraph II.B introduces the TL beam with transverse and longitudinal gradients that are exciting a magnetic quadrupole M2 transition. Corresponding multipole shifts are also evaluated. As a conclusion, a common "magic" trapping wavelength for a synthetic TL-induced optical frequency operational mode, based on a combination of E1M1 and M2 multipolar excitations, can be found by tuning the vector and tensor parts of the polarizability. All materials required for calculation are presented in the Appendix.

\section{Optical transitions induced by twisted-light (TL)}

\indent We provide now a detailed study of the resulting associated light shifts and magnetic sensibility which are dictated by the light beam decomposition into Bessel or Laguerre-Gauss modes near the center of the beam~\cite{Klimov:2012,Quinteiro:2017,Quinteiro:2023}.
\begin{table*}[t!!]
\begin{tabular}{c|cc|cc|c||c|c|c|c|c|c|c}
\hline
\hline
& $^{1}S_{0}~^{1}P_{1}$  & $^{1}P_{1}~^{3}P_{0}$ & $^{1}S_{0}~^{3}P_{1}$ & $^{3}P_{0}~^{3}P_{1}$ & $\lambda^{E1M1}_{\omega}$ & $A_{M2}$  & $^{1}S_{0}~^{3}P_{2}$ & $\Delta m=0$ & $\Delta m=\pm1$ & $\Delta m=\pm2$ & $\lambda^{M2}_{\omega}$ & \\
 & $\langle E1\rangle/ea_{0}$ & $\langle M1\rangle/\mu_{B}$ & $\langle E1\rangle/ea_{0}$ & $\langle M1\rangle/\mu_{B}$ & nm & mHz  & $\langle M2\rangle/\mu_{B}a_{0}$& MHz/T$^{2}$  & MHz/T$^{2}$ & Hz/T$^{2}$ & nm  & \\
\hline

$^{88}$Sr &  5.28~\cite{Porsev:2001} & 0.023~\cite{Santra:2005} & 0.15~\cite{Alden:2014} & $0.816$~\cite{Taichenachev:2006} & 1397 & 0.13~\cite{Derevianko:2001}  & 11  & 5.6 & 4.1 & & 671 & \\

$^{172}$Yb  &  4.40~\cite{Porsev:1999} & 0.103 & 0.54~\cite{Alden:2014} & $0.815$ & 1157 & 0.25~\cite{Flaumbaum:2018}  & 7.5  & 1.2~\cite{Flaumbaum:2018} & 0.92~\cite{Flaumbaum:2018} & -47~\cite{Flaumbaum:2018} & 507 & \\

$^{200}$Hg  &  2.80~\cite{Petersen:2008} & 0.140 & 0.46~\cite{Petersen:2008} & $0.804$ & 531 & 3.6~\cite{Garstang:1967}  & 3.8  & 0.45 & 0.34 & & 227 & \\

$^{24}$Mg  &  4.03~\cite{Porsev:2001}  & 0.063 & 0.0057~\cite{Alden:2014} & $0.814$ & 916 & 0.44~\cite{Derevianko:2001}  & 7.6 & 52.5 & 39.3 & & 456 & \\

$^{40}$Ca  &  4.91~\cite{Porsev:2001}  & 0.017 & 0.036~\cite{Alden:2014} & $0.816$ & 1319 & 0.13~\cite{Derevianko:2001}  & 10  & 20.6 & 15.4 & & 655 & \\

$^{112}$Cd  &  3.36~\cite{Yamaguchi:2019} & 0.036 & 0.15~\cite{Yamaguchi:2019} & $0.815$ & 664 & 0.96~\cite{Garstang:1967} & 10  & 1.8  & 1.4 & & 314 & \\

\hline
\hline
\end{tabular}
\caption{Reduced matrix elements for electric dipole E1, magnetic
dipole M1 and magnetic quadrupole M2 transitions. Einstein's coefficients $A$ for the quadrupole transition are given. Note that our $\langle$M1$\rangle/\mu_{B}$ reduced matrix element for the $^{1}$P$_{1}\rightarrow^{3}$P$_{0}$ transition in Sr is consistent with the theoretical value reported by~\cite{Santra:2005}. Second-order Zeeman-shifts of $\Delta\textup{m}$ sublevels of the $^{1}$S$_{0}\rightarrow^{3}$P$_{2}$ quadrupole components are given. Our evaluation of second-order Zeeman-shifts for $\Delta\textup{m}=0,\pm1$ components in Yb is consistent with the theoretical values reported in~\cite{Flaumbaum:2018}. Optical clock wavelengths $\lambda^{E1M1}_{\omega}$ and $\lambda^{M2}_{\omega}$ are indicated. (See Appendix for additional information).}
\label{table-I}
\end{table*}
For TL photons, the interplay  between orbital (OAM) and spin (SAM) angular momentum near the beam axis in the vicinity of the focusing spot leads to a strong modification of the local polarization and energy propagation~\cite{Monteiro:2009,Klimov:2012,Quinteiro:2017}. This unusual local polarization effect is indeed enhanced when orbital and spin momentum correspond to opposite helicities (i.e. when these angular momenta have an opposite projection along the propagation axis). Here we use this polarization and wave-structure in order to yield selection rules that open the forbidden transitions while maintaining light shifts at a level compatible with a very accurate clock operation.

\subsection{TL induced a two-photon E1M1 excitation}

\begin{figure}[t!!]
\center
\resizebox{8.5cm}{!}{\includegraphics[angle=0]{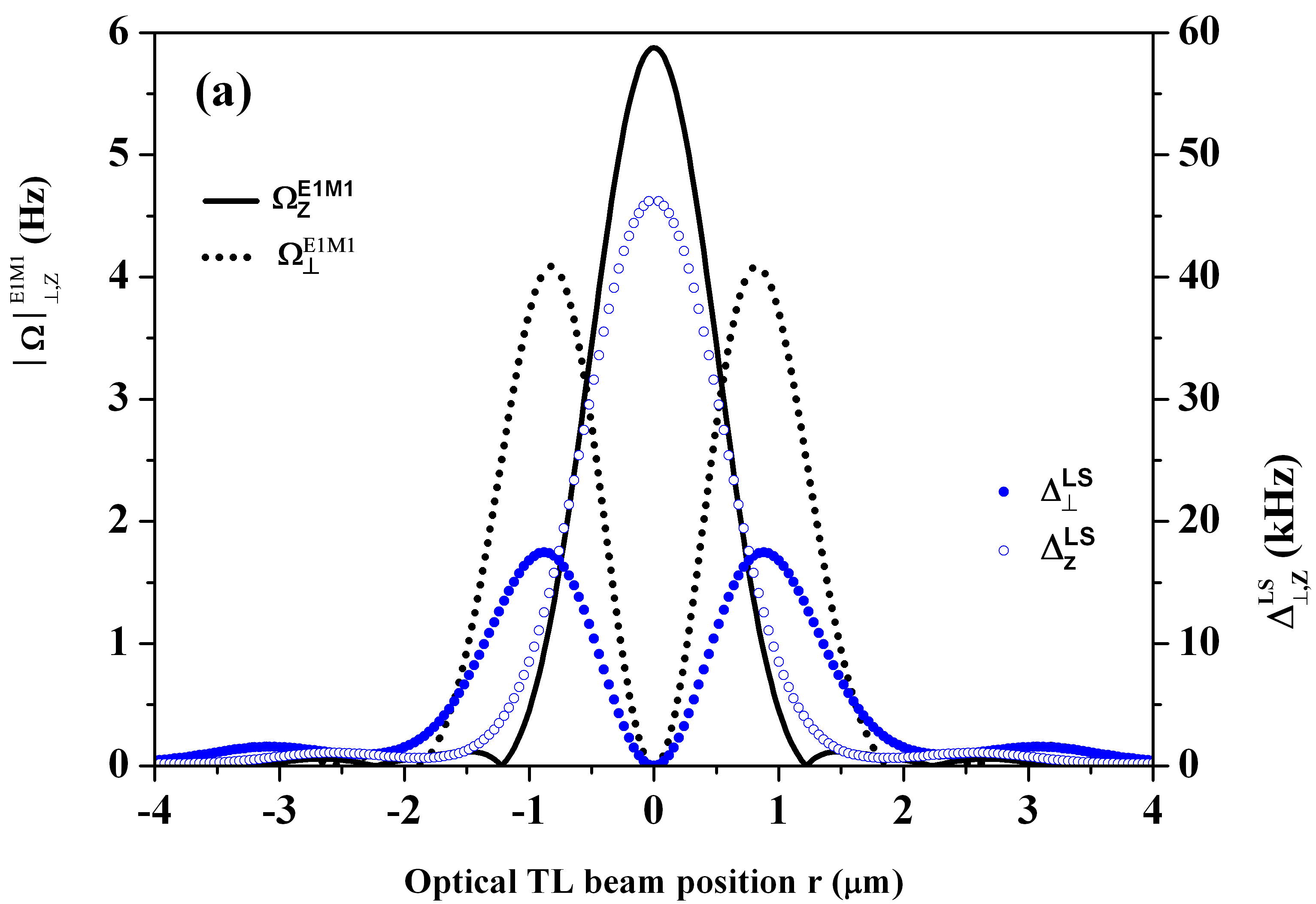}}
\resizebox{8.5cm}{!}{\includegraphics[angle=0]{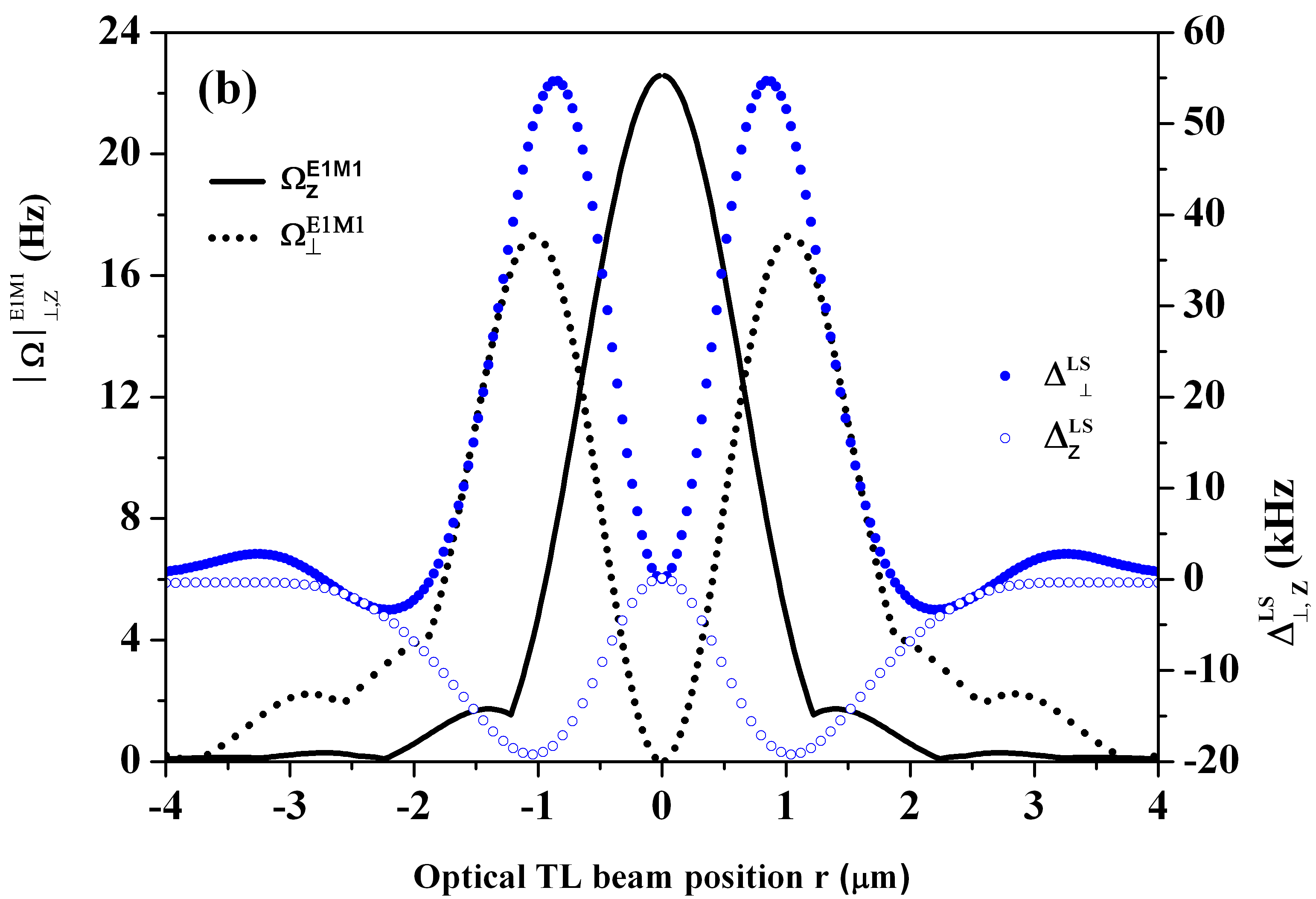}}
\caption{(color online). (a) E1M1 monochromatic two-photon excitation rate of the $^{88}$Sr two-photon transition at 1397~nm versus spatial position around the laser beam axis center. We have used OAM with $l=\pm1$ and SAM with $\sigma=\mp1$ (z: longitudinal component and $\perp$: transverse component). The associated light-shift of the clock transition induced by the monochromatic excitation is reported along the vertical right-axis. Laser power is $\textup{P}_{1397}\sim1$~mW with a waist of w$_{1397}\sim2~\mu$m. (b) E1M1 bichromatic excitation rate at 3236~nm and 890~nm versus spatial position around the laser beam axis center. The associated light-shift of the clock transition induced by the bichromatic excitation is reported along the vertical right-axis. Laser power are respectively $\textup{P}_{3236}\sim0.3$~mW ($\textup{P}_{890}\sim13$~mW) and a waist of w$_{3236}\sim4~\mu$m (w$_{890}\sim2~\mu$m). Curves are plotted for a well-localized atomic distribution in space. Note that typical peak values of longitudinal electric and magnetic oscillatory fields are respectively below 500~kV$/$m and 1~mT for the specific choice of laser parameters.}
\label{fig:E1M1}
\end{figure}

\indent To open the E1M1 two-photon transition and to excite the M2 magnetic quadrupole transition with TL as shown in Fig.~\ref{fig:fig1}, we need to evaluate coupling rates with appropriate expressions of electric and magnetic fields. The electric-dipole E1 and magnetic-dipole M1 elements required to excite the E1M1 clock transition are given by~\cite{Quinteiro:2017}:
\begin{equation}
\begin{split}
\langle^{1}S_{0}|H_{E1}|k\rangle_{\perp,z}&=-\langle E1\rangle\cdot\int_{0}^{1}\widetilde{E}_{\perp,z}(u\textbf{r})du\\
\langle^{3}P_{0}|H_{M1}|k\rangle_{\perp,z}&=-\langle M1\rangle\cdot\int_{0}^{1}u\widetilde{B}_{\perp,z}(u\textbf{r})du
\end{split}
\label{coupling-rate-E1M1}
\end{equation}
where intermediate atomic states are labeled by $|k\rangle\equiv|^{1}P_{1}\rangle,|^{3}P_{1}\rangle$.
We decompose the electric and magnetic fields by transverse $\perp$ and longitudinal $z$ components following~\cite{Quinteiro:2017}:
\begin{equation}
\begin{split}
\widetilde{\textbf{E}}(\textbf{r})&=\widetilde{E}_{\perp}(\textbf{r})\widehat{\textbf{r}}+\widetilde{E}_{z}(\textbf{r})\widehat{\textbf{z}}\\
\widetilde{\textbf{B}}(\textbf{r})&=\widetilde{B}_{\perp}(\textbf{r})\widehat{\textbf{r}}+\widetilde{B}_{z}(\textbf{r})\widehat{\textbf{z}}
\end{split}
\label{electromagnetic-fields}
\end{equation}
where the expressions of $\widetilde{E}_{\perp}(\textbf{r}),\widetilde{B}_{\perp}(\textbf{r})$ and $\widetilde{E}_{z}(\textbf{r}),\widetilde{B}_{z}(\textbf{r})$ are given in the Appendix A1.
The non degenerate two-photon E1M1 transition using a bichromatic angular frequency is~\cite{Goppert-Mayer:1931,Alden:2014,Jackson:2019}:
\begin{equation}
\begin{split}
\Omega^{E1M1}_{\perp,z}(\textbf{r})=\left|\sum_{k}\sum_{i\neq j}\frac{\langle^{1}S_{0}|H^{i}_{E1}|k\rangle_{\perp,z}\langle k|H^{j}_{M1}|^{3}P_{0}\rangle_{\perp,z}}{\hbar\omega_{i}-E(k)}\right|
\end{split}
\label{two-photon-rate}
\end{equation}
where we have defined $E(^{1}S_{0})=0$. One sum is over intermediate non resonant atomic state $|k\rangle$ and the second one is for each off degenerate $\omega_{i}$ photon frequency. For a monochromatic excitation, the summation over $i\neq j$ in Eq.~\ref{two-photon-rate} is not applied (see Appendix A2). For the bichromatic two-photon E1M1 transition driven by $\omega_{1}$ and $\omega_{2}$ photons, Eq.~\ref{two-photon-rate} includes two separate contributions with the permutation of $\omega_{1},\omega_{2}$ indices into electric and magnetic fields.
Reduced $\langle E1\rangle$ and $\langle M1\rangle$ matrix elements are reported into Table~\ref{table-I} for different atoms using Mizuchima formulae given in Appendix A3.\\
\indent We address the problem of the light-shift associated to the E1M1 clock excitation. The dominant contribution is coming from electric-dipole scalar terms that are effectively shifting the $^{1}$S$_{0}$ and $^{3}$P$_{0}$ atomic states. We have reported the evaluation of the E1M1 transition rate and the light-shift effect in Fig.~\ref{fig:E1M1}(a) for a $^{88}$Sr atom localized at the center of the beam. As expected, electric and magnetic longitudinal field components $\widetilde{E}_{z}(\textbf{r}),\widetilde{B}_{z}(\textbf{r})$ are producing a large two-photon transition rate near the beam center while the transverse $\widetilde{E}_{\perp}(\textbf{r}),\widetilde{B}_{\perp}(\textbf{r})$ electromagnetic fields are vanishing.
In experimental situations, atoms are not anymore well localized spatially and explore a few hundreds of nanometers in the transverse and longitudinal direction.
Assuming a realistic scenario, the transition rate given by Eq.~\ref{two-photon-rate} is weighted by a Gaussian spatial distribution of the atomic sample in transverse and longitudinal directions.
Averaging transition rate and light-shift contributions lead to corrected expressions as following:
\begin{equation}
\begin{split}
\overline{\Omega}^{E1M1}_{\perp,z}(\textbf{r}_{m})&=\frac{1}{2\pi\rho_{\perp,z}^{2}}\int\Omega^{E1M1}_{\perp,z}(\textbf{r})e^{-\frac{(\textbf{r}-\textbf{r}_{m})^{2}}{2\rho_{\perp,z}^{2}}}d^{2}\textbf{r}\\
\overline{\Delta}^{LS}_{\perp,z}(\textbf{r}_{m})&=\frac{1}{2\pi\rho_{\perp,z}^{2}}\int\Delta^{LS}_{\perp,z}(\textbf{r})e^{-\frac{(\textbf{r}-\textbf{r}_{m})^{2}}{2\rho_{\perp,z}^{2}}}d^{2}\textbf{r}
\end{split}
\label{weighted-two-photon-rate}
\end{equation}
where $\textbf{r}_{m}$ is the most probable impact parameter and $\rho_{\perp,z}$ is the width of the atomic distribution. The on-axis and off-axis atomic excitations by twisted light as a function of the impact parameter, i.e., the distance from the \textbf{r} position, was investigated in~\cite{Afanasev:2013}. An accurate analysis with an integration over the impact parameter was presented in~\cite{Scholz-Marggraf:2014}.
The spatial distribution of the excitation rate and light-shift is related to the spatial shape distribution of transversal $\widetilde{E}_{\perp}(\textbf{r})$ and longitudinal $\widetilde{E}_{z}(\textbf{r})$ electric fields. The average process at $\textbf{r}_{m}=0$ gives a corrected longitudinal two-photon Rabi field excitation of 4~Hz and a transverse contribution of 2~Hz setting a width $\rho_{\perp,z}=300$~nm with a laser power of P$_{1397}=1$~mW and a waist of w$_{1397}=2$~$\mu$m.
The average longitudinal light-shift $\overline{\Delta}^{LS}_{z}(0)$ of the clock transition is found to be 35~kHz while the transverse contribution $\overline{\Delta}^{LS}_{\perp}(0)$ reaches 7~kHz.
\begin{table}[t!!]
\begin{tabular}{c|c|c|cc|cc}
\hline
\hline
$\lambda$ & w$_{\lambda}$ & P$_{\lambda}$  & $\overline{\Omega}^{E1M1}_{z}(0)$ & $\overline{\Omega}^{E1M1}_{\perp}(0)$ & $\overline{\Delta}^{LS}_{z}(0)$ & $\overline{\Delta}^{LS}_{\perp}(0)$  \\
nm & $\mu$m & mW & Hz & Hz & kHz & kHz \\
\hline
$\begin{array}{c}
  890 \\
  3236
\end{array}$
  & $\begin{array}{c}
  2 \\
  4
\end{array}$ &  $\begin{array}{c}
  13 \\
  0.3
\end{array}$  & 18 & 5.6 & $\begin{array}{c}
  +30 \\
  -52
\end{array}$ & $\begin{array}{c}
  +23.4 \\
  -1.4
\end{array}$  \\
\hline
$\begin{array}{c}
  441 \\
  1200
\end{array}$  & $\begin{array}{c}
  2 \\
  2
\end{array}$ & $\begin{array}{c}
  0.35 \\
  2
\end{array}$  & 7 & 12.5 & $\begin{array}{c}
  -8 \\
  +30
\end{array}$ & $\begin{array}{c}
  -32 \\
  +10
\end{array}$  \\
\hline
$\begin{array}{c}
  375.6 \\
  813.4
\end{array}$ & $\begin{array}{c}
  2 \\
  2
\end{array}$ & $\begin{array}{c}
  0.1 \\
  20
\end{array}$ & 2 & 6.3 & $\begin{array}{c}
  -0.25 \\
  0
\end{array}$ & $\begin{array}{c}
  -1.3 \\
  0
\end{array}$  \\
\hline
\hline
\end{tabular}
\caption{Three possible sets of bichromatic E1M1 laser excitation parameters with significant excitation rates to reduce or compensate for the total light-shift affecting the $^{88}$Sr clock transition with delocalized atoms around $\textbf{r}_{m}=0$ described by a typical spatial width $\rho_{\perp,z}=300$~nm.}
\label{table-II}
\end{table}
Those light-shifts are due to the presence of strong electric fields shown on in Fig.~\ref{fig:E1M1}(a) following the vertical right-axis. However, there is some experimental flexibility by playing simultaneously with the waist size or the laser intensity to reduce the light-shift below a few kHz while keeping a two-photon excitation rate at the Hz level.\\
\indent The previous method can be extended to a bichromatic two-photon twisted-light scheme to efficiently compensate for the overall light-shift affecting the clock frequency measurement. Complete elimination of the two-photon light-shift during a E1M1 excitation is possible with dual frequencies~\cite{Alden:2014}. We have reported the bichromatic excitation rate and the light-shift compensation in Fig.~\ref{fig:E1M1}(b) using the first set of two different wavelengths and TL intensity given in Table~\ref{table-II}. After averaging over a spatial delocalization $\rho_{\perp,z}=300$~nm, total transverse and longitudinal light-shifts $\overline{\Delta}^{LS}_{\perp,z}(0)\sim\pm22$~kHz can mutually compensate quite efficiently.

Non-degenerated two photon E1M1 excitation, where one of the laser is blue detuned with respect to the main electric dipole transition, might facilitate the practical implementation of the scheme. Indeed, a blue detuned TL beam offers a transversally stable trapping condition which also secure that a single trapped atom will be located at the centre of the TL beam, minimizing systematic errors due to beam misalignments~\cite{Xu:2010}. As an illustrative example, we take the parameters values of both lasers in the second line in Table~\ref{table-II} and we find a trap height around $21\,\mu$K in temperature unit with a trapping frequency around $7\,$kHz (done in the paraxial approximation), which are small values but still suitable for optical tweezers operation with Sr atoms cooled on the intercombination line~\cite{Chalony:2011}. Signal to noise ratio can be drastically increased considering few hundreds of atoms trapped in individual TL optical tweezers using recently developed tweezers-array techniques~\cite{Barredo:2016,Young:2020,Kaufman:2021,Tian:2023}. A third set of parameters from Table~\ref{table-II} can be used where one of the laser light is at the metrological magic wavelength to trap atoms~\cite{Derevianko:2011} with zero light-shift while the second laser excitation wavelength has a small contribution to the overall light-shift that can be easily reduced to a vanishing correction by auto-balanced Ramsey spectroscopy~\cite{Sanner:2018,Yudin:2018}.\\
\begin{figure}[b!!]
\center
\resizebox{8.5cm}{!}{\includegraphics[angle=0]{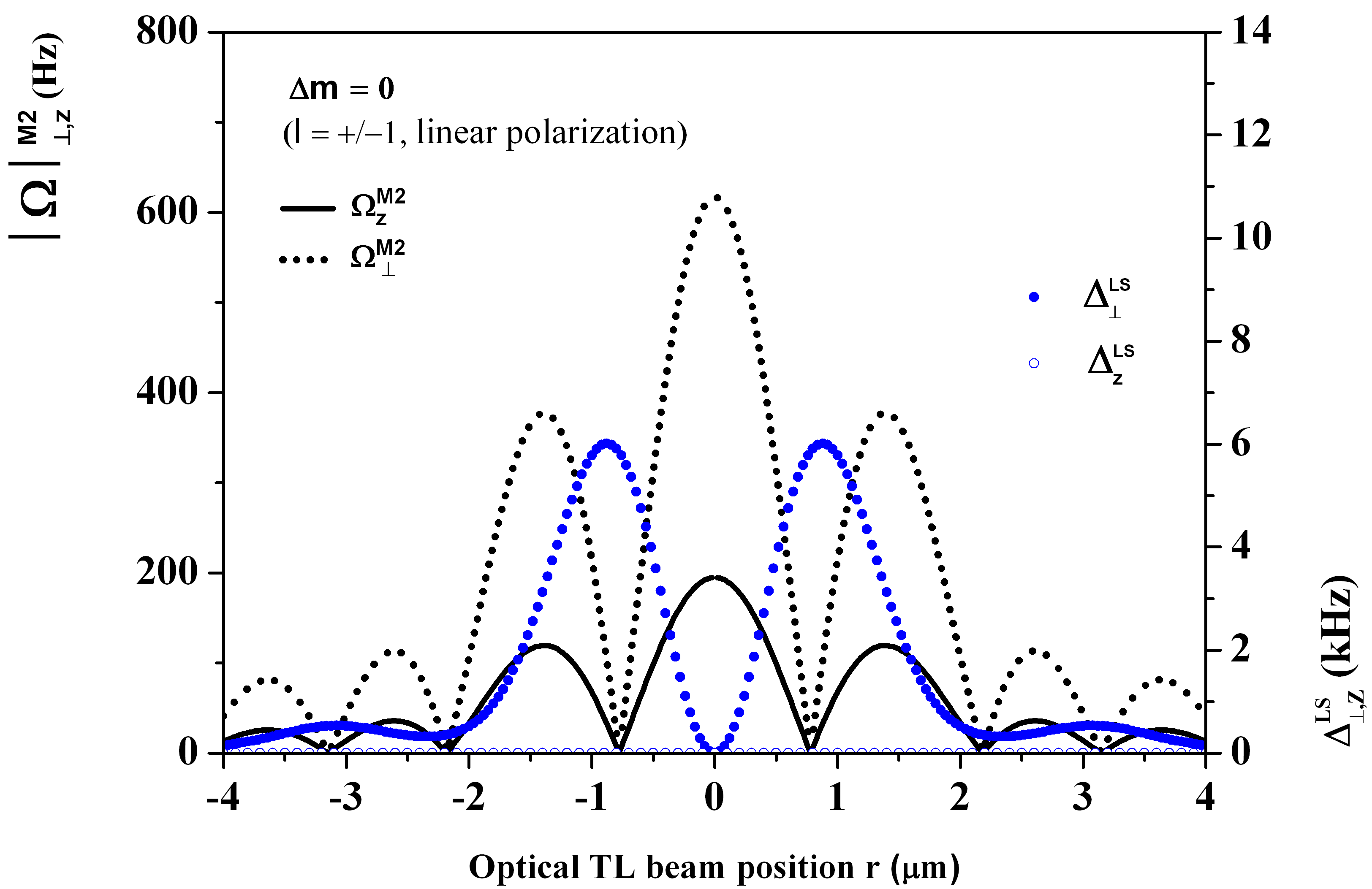}}
\caption{(color online). Excitation rate of the Zeeman magnetic-insensitive M2 quadrupole clock transition of the $^{88}$Sr induced by a TL beam at 671~nm. The associated light-shift of the clock transition is reported along the vertical right-axis. Laser power is fixed to $\textup{P}_{671}\sim1$~mW with a waist of w$_{671}\sim2~\mu$m.}
\label{fig:M2}
\end{figure}
\subsection{TL induced a single-photon M2 excitation}

\indent Turning to the second narrow clock transition shown in Fig.~\ref{fig:fig1}, the excitation rate by a TL beam of the $^{1}$S$_{0}\rightarrow^{3}$P$_{2}$ magnetic quadrupole can be simply approximated as:
\begin{equation}
\begin{split}
\Omega^{M2}_{\perp,z}(\textbf{r})=-\frac{1}{6}\langle M2\rangle\cdot\nabla_{\textbf{r}}\int_{0}^{1}u\widetilde{B}_{\perp,z}(u\textbf{r})du
\end{split}
\label{M2-quadrupole-rate}
\end{equation}
where the quadrupole tensor component has been replaced by a reduced matrix element $\langle M2\rangle$. Coupling elements $\langle M2\rangle$ are derived from Einstein's coefficients reported in Table~\ref{table-I} for different atoms with first-order and second-order Zeeman shifts estimated within Appendix A4.
\begin{figure}[t!!]
\center
\resizebox{8.5cm}{!}{\includegraphics[angle=0]{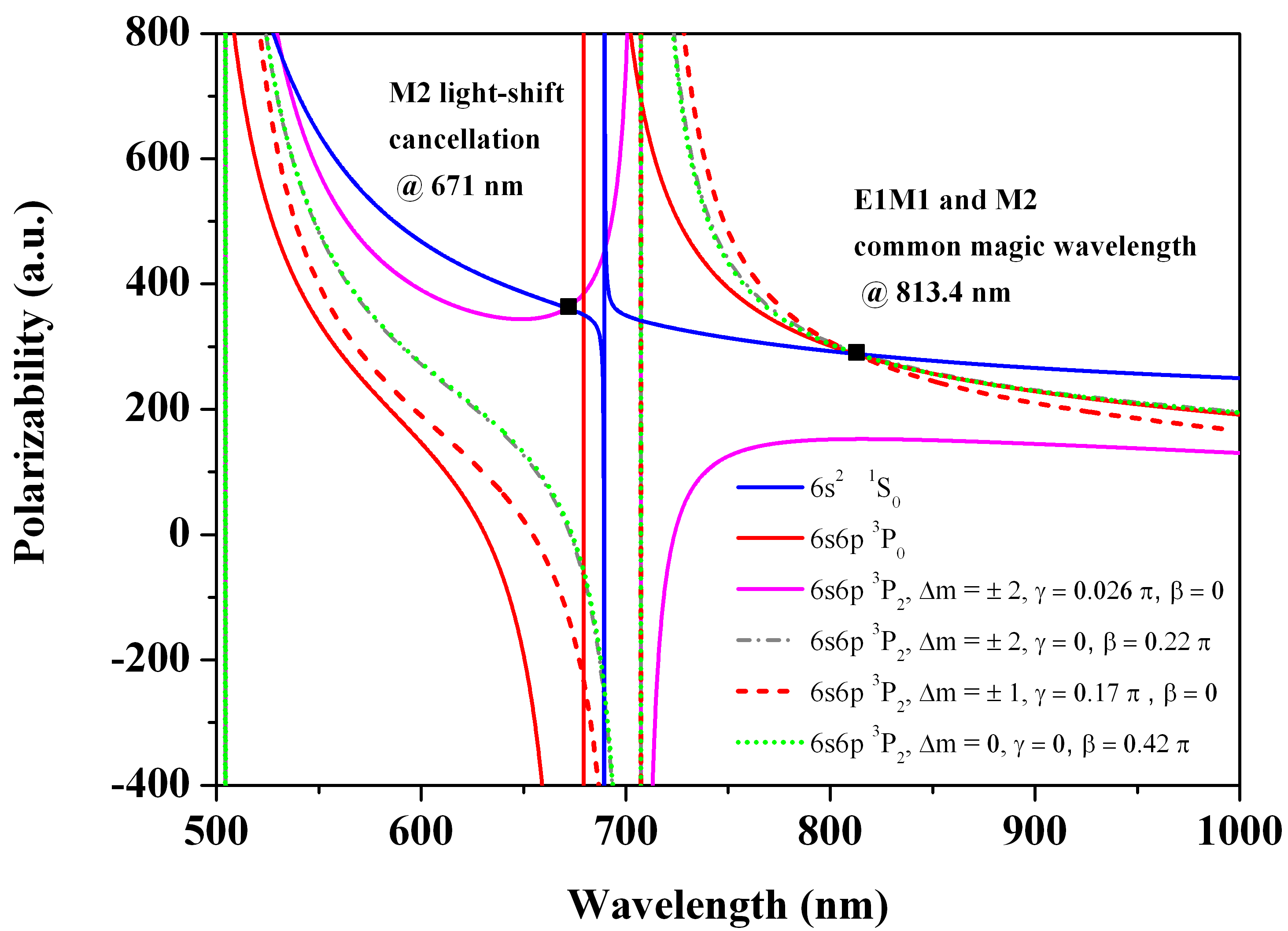}}
\caption{(color online). Tunability of $^{88}$Sr atomic state polarizabilities of the $^{1}$S$_{0}\rightarrow^{3}$P$_{0}$ and $^{1}$S$_{0}\rightarrow^{3}$P$_{2}$ clock transitions including Zeeman components $\Delta\textup{m}$ for different trapping wavelengths under various sets of polarization angle ellipticity $\gamma$ and projection angle $\beta$ as defined in~\cite{Trautmann:2023}. We have used 1 a.u.$=4\pi\epsilon_{0}a_{0}^{3}$ where $\epsilon_{0}$ is the vacuum permittivity and $a_{0}$ the Bohr radius. Magic wavelengths are marked by black squares $\blacksquare$. Only the $\Delta\textup{m}=2$ light-shift of the quadrupole transition can be eliminated by a magic polarization angle ellipticity $\gamma=0.026\pi$ at the clock wavelength excitation of 671~nm. Note that grey short dashed dots and green short dots are nearly indistinguishable.}
\label{fig:E1M1-M2-LS}
\end{figure}
The vectorial and tensorial nature of the quadrupole transition makes possible to selectively excite all Zeeman components while compensating the light-shift not only by circular or linear polarized TL beam modes~\cite{Schmiegelow:2012,Schmiegelow:2016,Quinteiro:2017-bis} but also through the tunability of the atomic state polarizability~\cite{Trautmann:2023,Ishiyama:2023,Tang:2023}. Note that Zeeman components of a quadrupole transition are driven by TL excitation and the angular dependence of the vector spherical harmonics related to the quantization axis orientation~\cite{Schmiegelow:2012,Schulz:2020}.
The magnetic-insensitive Zeeman $\Delta\textup{m}=0$ component, reported in Fig.~\ref{fig:M2}, can be used as a clock transition driven by a linear polarized TL vortex beam with a Laguerre-Gauss (LG) or a Bessel mode where electric fields are vanishing in the beam axis center~\cite{Schmiegelow:2012}. If desired, magnetic sensitive $\Delta\textup{m}=\pm1$ ($\Delta\textup{m}=\pm2$) components are also efficiently excited by a combination of OAM $l=\pm2$ ($l=\pm1$) and SAM $\sigma=\mp1$ ($\sigma=\pm1$) still with vanishing electric fields (see also Appendix A5 and A6). Residual electromagnetic fields and gradients lead to multipole shifts (see Appendix A7) below 200~mHz due to off-resonant M1, E2 and M3 decay channels~\cite{Flaumbaum:2018,Derevianko:2001}, all synchronized with TL excitation pulses and thus easily removed by composite pulse spectroscopy~\cite{Zanon-Willette:2018}.
In addition to our innovative approach, achieving a triple-state magic trapping condition at the metrological 813.4~nm wavelength for all Zeeman components is realized by tuning either the vector or the tensor contribution through polarization ellipticity of the trapping light or by suitably adjusting polarization angles along the quantization axis as recently demonstrated in Sr~\cite{Trautmann:2023} and in Yb~\cite{Tang:2023}. Our TL scheme based on the M2 quadrupole transition benefits from the additional tunability of the vector polarizability realizing suppression of ac Stark-shifts for $\Delta\textup{m}=\pm2$ sublevels at the 671~nm clock frequency when the magic ellipticity $\gamma=0.026\pi$ defined by~\cite{Trautmann:2023} is reached.
These additional control knobs and magic angle conditions for the magnetic quadrupole components of $^{88}$Sr are reported in Fig.~\ref{fig:E1M1-M2-LS} based on the material presented in Appendix A5.
Finally, the second-order Zeeman-shift, given in Table~\ref{table-I}, can exhibit a magnitude ten times smaller than the conventional $^{1}$S$_{0}\rightarrow^{3}$P$_{0}$ fermionic or bosonic clock transition for some atomic species~\cite{Flaumbaum:2018,Safronova:2018,Ishiyama:2023}.

\section{Conclusion}

\indent Having a set of two clock interrogation schemes with different sensitivity to external fields will help to reduce systematic uncertainties~\cite{Safronova:2018,Bohman:2023}, for instance in the synthetic frequency operational mode~\cite{Yudin:2011,Yudin:2021}. This interrogation mode will be supported by a bi-color resonance excitation at the same magic wavelength simplifying interleaved sequential clock operations to probe common environmental distortions~\cite{Safronova:2018}.
\indent In conclusion, we propose an all-optical spectroscopic method replacing the plane-wave interaction by polychromatic TL beams probing ultra-narrow atomic resonances with multiphoton excitations. Our proposed scheme represents an innovative way based on the use of two different atomic transitions in the same quantum system and well controlled parameters of the exciting TL laser. Beyond frequency metrology, forbidden transitions are natural candidates as resilient qubits for quantum information processing due to their large coherence time. The integration of TL beams and forbidden transitions into quantum computing architectures such as scalable arrays of optical tweezers represents a valuable technique for optical clocks, quantum computation and simulation~\cite{Barredo:2016,Young:2020,Kaufman:2021,Tian:2023}. Matter-wave interferometry can be envisaged manipulating quantum interferences by TL beams and spatially movable optical tweezer traps~\cite{Nemirovsky:2023,Premawardhana:2023}.

\indent T.Z.W acknowledges E. Peik and A. Surzhykov for comments, J. Trautmann for providing recent atomic state polarizabilities of Sr and L. Pruvost for interest in this work. F.I. was supported by the Brazilian agencies CNPq (310265/2020-7), CAPES and FAPERJ (210.296/2019). This work was partially supported by INCT-IQ (465469/2014-0) and by the National research foundation and Quantum Engineering Programme No. NRF2021-QEP2-03-P01. V.I.Yudin was supported by the Russian Science Foundation (project no. 21-12-00057), and by the Ministry of Science and Higher Education of the Russian Federation (project no. FSUS-2020-0036).
A.V. Taichenachev acknowledges financial support from Russian Science Foundation through the grant 20-12-00081.

\section*{Appendix}

\section*{A1. Electric and magnetic fields of twisted-light beams}

\indent Twisted-light (TL) fields are characterized by the orbital angular momentum (OAM) called a topological charge $l$ which is adding to the spin angular momentum (SAM) defined by the helicity (polarization) $\sigma$  of the laser beam. Electric and magnetic components that are propagating along the Oz axis are introduced as following~\cite{Quinteiro:2015,Quinteiro:2017}:
\begin{equation}
\begin{split}
\textbf{E}(\textbf{r},t)&=\frac{1}{2}\widetilde{\textbf{E}}(\textbf{r})e^{i(q_{z}z-\omega t)}+c.c.\\
\textbf{B}(\textbf{r},t)&=\frac{1}{2}\widetilde{\textbf{B}}(\textbf{r})e^{i(q_{z}z-\omega t)}+c.c.
\end{split}
\label{propagating-waves}
\end{equation}
where c.c means complex conjugate.
Electric and magnetic fields are decomposed into a transverse $\perp$ component and a longitudinal $z$ component as~\cite{Quinteiro:2015,Quinteiro:2017}:
\begin{equation}
\begin{split}
\widetilde{\textbf{E}}(\textbf{r})&=\widetilde{E}_{\perp}(\textbf{r})\widehat{\textbf{r}}+\widetilde{E}_{z}(\textbf{r})\widehat{\textbf{z}}\\
\widetilde{\textbf{B}}(\textbf{r})&=\widetilde{B}_{\perp}(\textbf{r})\widehat{\textbf{r}}+\widetilde{B}_{z}(\textbf{r})\widehat{\textbf{z}}
\end{split}
\label{electromagnetic-fields}
\end{equation}
where $\widetilde{E}_{\perp}(\textbf{r})\widehat{\textbf{r}}=\widetilde{E}_{x}(\textbf{r})\widehat{\textbf{x}}+\widetilde{E}_{y}(\textbf{r})\widehat{\textbf{y}}$ and $\widetilde{B}_{\perp}(\textbf{r})\widehat{\textbf{r}}=\widetilde{B}_{x}(\textbf{r})\widehat{\textbf{x}}+\widetilde{B}_{y}(\textbf{r})\widehat{\textbf{y}}$.

Non paraxial expressions of transverse and longitudinal electromagnetic fields of the twisted-light are for Bessel beams given by~\cite{Quinteiro:2017}:
\begin{equation}
\begin{split}
\widetilde{E}_{x}(\textbf{r})=&iE_{0}J_{l}\left(q_{r}r\right)e^{il\varphi}\\
\widetilde{E}_{y}(\textbf{r})=&-\sigma~E_{0}J_{l}\left(q_{r}r\right)e^{il\varphi}\\
\widetilde{B}_{x}(\textbf{r})=&\sigma~B_{0}\left[\left(1+\frac{q_{r}^{2}}{2q^{2}_{z}}-\frac{q_{r}^{2}}{2q^{2}_{z}}e^{i2\sigma\varphi}\right)J_{l}\left(q_{r}r\right)e^{il\varphi}\right.\\
&+\left.\frac{q_{r}^{2}}{2q^{2}_{z}}\left(l+\sigma\right)\frac{2}{q_{r}r}J_{l+\sigma}\left(q_{r}r\right)e^{i(l+2\sigma)\varphi}\right]\\
\widetilde{B}_{y}(\textbf{r})=&iB_{0}\left[\left(1+\frac{q_{r}^{2}}{2q^{2}_{z}}+\frac{q_{r}^{2}}{2q^{2}_{z}}e^{i2\sigma\varphi}\right)J_{l}\left(q_{r}r\right)e^{il\varphi}\right.\\
&-\left.\frac{q_{r}^{2}}{2q^{2}_{z}}\left(l+\sigma\right)\frac{2}{q_{r}r}J_{l+\sigma}\left(q_{r}r\right)e^{i(l+2\sigma)\varphi}\right]\\
\widetilde{E}_{z}(\textbf{r})=&\sigma E_{0}\frac{q_{r}}{q_{z}}J_{l+\sigma}\left(q_{r}r\right)e^{i(l+\sigma)\varphi}\\
\widetilde{B}_{z}(\textbf{r})=&-iB_{0}\frac{q_{r}}{q_{z}}J_{l+\sigma}\left(q_{r}r\right)e^{i(l+\sigma)\varphi}
\label{electromagnetic-field}
\end{split}
\end{equation}
where $q_{r}^{2}+q_{z}^{2}=(n\omega/c)^{2}$, $1/q_{r}$ is a measure of the beam radius defined as $q_{r}=\frac{2\pi}{w}$, the waist $w$, $n$ being the refractive index of the medium, $\omega$ is the angular frequency, $B_{0}=q_{z}E_{0}/\omega$, $r=\sqrt{x^{2}+y^{2}}$ and $\varphi=\arctan[y/x]$.

We note that TL or vortex beams are ususally decomposed over Laguerre-Gauss modes including the general case of elliptic polarization~\cite{Klimov:2012,Koksal:2022}.

\section*{A2. Monochromatic two-photon transition rate}

\indent We derive the monochromatic two-photon expression used in the main text to evaluate the E1M1 excitation rate in Hz. By repeating the
treatments of~\cite{Goppert-Mayer:1931} and~\cite{Jackson:2019}, the admixture for the wave-function from the oscillatory field at angular frequency $\omega=E(^{3}P_{0})/2\hbar$ is given by:
\begin{equation}
\begin{split}
|^{3}P'_{0}\rangle\approx|^{3}P_{0}\rangle&+\sum_{k}\frac{\langle k| H_{M1}|^{3}P_{0}\rangle}{\Delta_{k}-\hbar\omega}|k\rangle\\
\approx|^{3}P_{0}\rangle&+\frac{\langle^{1}P_{1}|H_{M1}|^{3}P_{0}\rangle}{\Delta_{1}-\hbar\omega}|^{1}P_{1}\rangle\\
&+\frac{\langle^{3}P_{1}|H_{M1}|^{3}P_{0}\rangle}{\Delta_{2}-\hbar\omega}|^{3}P_{1}\rangle\\
\end{split}
\label{E1M1-wave-function}
\end{equation}
Then the two-photon excitation rate expression is:
\begin{equation}
\begin{split}
\langle^{1}S_{0}|H_{E1}|^{3}P'_{0}\rangle\approx&\frac{\langle^{1}S_{0}|H_{E1}|^{1}P_{1}\rangle\langle^{1}P_{1}|H_{M1}|^{3}P_{0}\rangle}{\Delta_{1}-\hbar\omega}\\
&+\frac{\langle^{1}S_{0}|H_{E1}|^{3}P_{1}\rangle\langle^{3}P_{1}|H_{M1}|^{3}P_{0}\rangle}{\Delta_{2}-\hbar\omega}
\end{split}
\label{E1M1-rate}
\end{equation}
with $\Delta_{1}=E(^{3}P_{0})-E(^{1}P_{1})$ and $\Delta_{2}=E(^{3}P_{0})-E(^{3}P_{1})$.
The monochromatic two-photon transition rate is then~\cite{Jackson:2019}:
\begin{equation}
\begin{split}
\Omega^{E1M1}_{\perp,z}=\sum_{k}\frac{\langle^{1}S_{0}|H_{E1}|k\rangle_{\perp,z}\langle k|H_{M1}|^{3}P_{0}\rangle_{\perp,z}}{\hbar\omega-E(k)}
\end{split}
\label{Rabi-rate}
\end{equation}
where $|k\rangle\equiv|^{1}P_{1}\rangle,|^{3}P_{1}\rangle$. The bichromatic version of the two-photon excitation rate is given in the main text. In addition, a trichromatic version of a three-photon excitation rate expression can be elaborated and theoretically derived in~\cite{Hobson:2016}.
\begin{figure}[t!!]
\center
\resizebox{8.5cm}{!}{\includegraphics[angle=0]{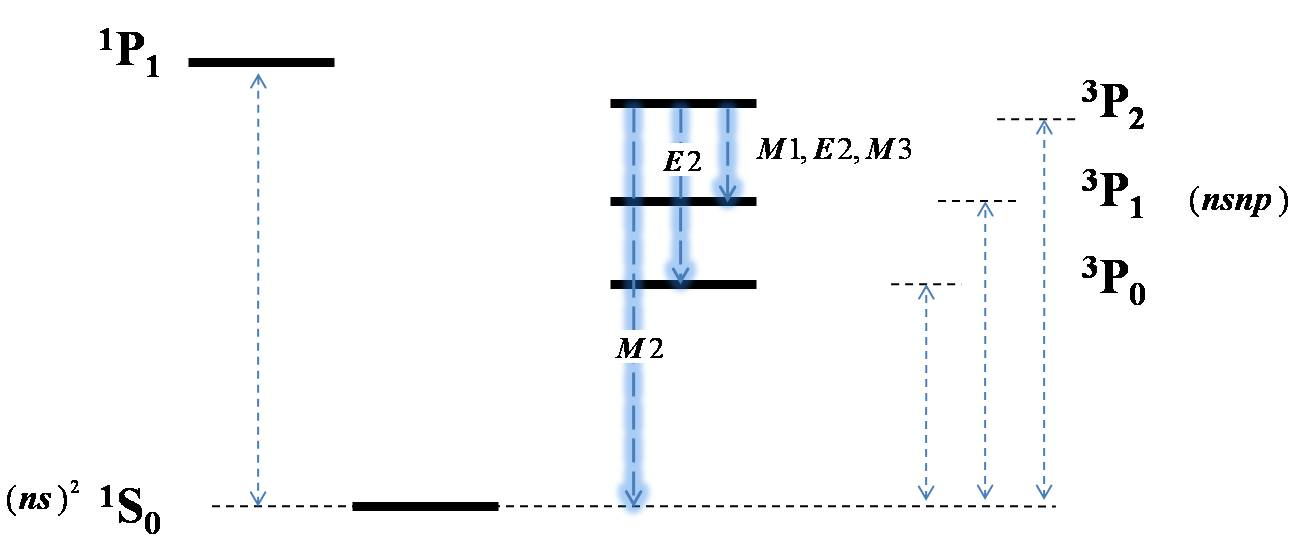}}
\caption{(color online). Single-photon decay channels from the $^{3}$P$_{2}$ state of alkaline-earth atoms~\cite{Derevianko:2001}.}
\label{fig:decay-channels}
\end{figure}
\begin{table*}[t!!]
\begin{tabular}{c|c|c|c||c|c|c||c|c|c|c|c}
\hline
\hline
& $A_{M2}$   & $^{3}P_{2}\rightarrow^{1}S_{0}~$ & $\lambda^{M2}_{\omega}$ & $A_{E2}$ & $^{3}P_{2}\rightarrow^{3}P_{0}$ & $\lambda^{E2}_{\omega}$ & $A_{E2}$ & $^{3}P_{2}\rightarrow^{3}P_{1}~$ & $A_{M1}$ & $^{3}P_{2}\rightarrow^{3}P_{1}~$ & $\lambda^{E2,M1}_{\omega}$ \\
& $\times10^{-3}s^{-1}$ & $\langle M2\rangle/\mu_{B}a_{0}$ & nm &  $\times10^{-6}s^{-1}$ & $\langle E2\rangle/ea^{2}_{0}$  & $\mu$m & $\times10^{-6}s^{-1}$ & $\langle E2\rangle/ea^{2}_{0}$  & $\times10^{-3}s^{-1}$ & $\langle M1\rangle/\mu_{B}$  & $\mu$m \\
\hline
$^{88}$Sr   & 0.13~\cite{Derevianko:2001}  & 11 & 671 & 1~\cite{Derevianko:2001} & 16.7 & 17.3 & 0.3~\cite{Derevianko:2001} & 24.4  & 0.83~\cite{Derevianko:2001} & $\sim0.7$ & 25.7 \\

$^{172}$Yb   & 0.25~\cite{Flaumbaum:2018}  & 7.5 & 507 &  &  & 4.1 & &  & 67~\cite{Flaumbaum:2018} & $\sim0.7$ & 5.8 \\

$^{200}$Hg   & 3.6~\cite{Garstang:1967}  & 3.8  & 227 & & & 1.56 & &  & & & 2.15 \\

$^{24}$Mg  & 0.44~\cite{Derevianko:2001}  & 7.6 & 456 & $3\times10^{-6}$~\cite{Derevianko:2001} & 7.9 & 163 & $1\times10^{-6}$~\cite{Derevianko:2001} & 12.2  & $9.12\times10^{-4}$~\cite{Derevianko:2001} & $\sim0.7$ & 242 \\

$^{40}$Ca  & 0.13~\cite{Derevianko:2001}  & 10 & 655 & $10^{-3}$~\cite{Derevianko:2001} & 13.4 & 63.3 & $3\times10^{-4}$~\cite{Derevianko:2001} & 20  & $1.6\times10^{-2}$~\cite{Derevianko:2001} & $\sim0.7$ & 94.4 \\

$^{112}$Cd  & 0.96~\cite{Garstang:1967} & 10  & 314 &  &  & 5.8 & &  & &  & 8.5 \\

\hline
\hline
\end{tabular}
\caption{Estimation of reduced $\langle M1\rangle,\langle M2\rangle$ and $\langle E2\rangle$ matrix elements by using Eq.~\ref{Mizuchima-E} and Eq.~\ref{Mizuchima-M} related to M1 and E2 decay channels with Einstein's coefficients reported from several references. Note that Sr, Mg and Ca decay channels of the $^{3}$P$_{2}$ state have been also evaluated by~\cite{Santra:2004}. Recent experimental values for Mg have been reported in~\cite{Jensen:2011} and for Sr in~\cite{Lu:2022}.}
\label{table-III}
\end{table*}

\section*{A3. Einstein coefficients and multipole moment evaluation}

\indent Electric $\langle Ej\rangle$ and magnetic $\langle Mj\rangle$ multipole reduced matrix elements are required with Einstein coefficients A$_{E_{j},M_{j}}$ ($j=1,2$) to evaluate Rabi excitation rates associated to dipole or quadrupole transitions in various atomic species.
We apply Mizuchima expressions that are for the electric multipole excitation given by~\cite{Mizushima:1964}:
\begin{equation}
\begin{split}
\langle Ej\rangle=\frac{1}{e a^{j}_{0}}\sqrt{A_{E_{j}}\frac{(2j+1)\{(2j-1)!\}^{2}(4\pi\hbar\epsilon_{0}c^{2j+1})}{j(j+1)\omega^{2j+1}}}
\end{split}
\label{Mizuchima-E}
\end{equation}
and for the magnetic multipole excitation given by~\cite{Mizushima:1966}:
\begin{equation}
\begin{split}
\langle Mj\rangle=\frac{1}{\mu_{B}a^{j-1}_{0}}\sqrt{A_{M_{j}}\frac{j\Gamma(j+3/2)(4\pi\hbar\epsilon_{0}c^{2j+3})}{(j+1)(2j+1)\omega^{2j+1}}}
\end{split}
\label{Mizuchima-M}
\end{equation}
where $\Gamma(j+3/2)$ is the Gamma function.
Angular frequency of the transition is $\omega$, the Bohr magneton $\mu_{B}$, the Bohr radius $a_{0}$, the electric charge $e$ and the vacuum permittivity $\epsilon_{0}$.
We have evaluated with Eq.~\ref{Mizuchima-E} and Eq.~\ref{Mizuchima-M} not only the magnetic quadrupole coupling element $\langle M2\rangle$ of the $^{1}$S$_{0}\rightarrow^{3}$P$_{2}$ clock transition but possible single photon decay channels for various atomic species~\cite{Derevianko:2001} as shown in Fig.~\ref{fig:decay-channels}. We have reported the resulting quantities into Table~\ref{table-I} and Table~\ref{table-III}.
\begin{table}[t!!]
\begin{tabular}{c|cc}
\hline
\hline
Atomic species & $\alpha,\beta$ \\
\hline

$^{88}$Sr  & $0.9996,-0.0285$~\cite{Boyd:2007} \\

$^{172}$Yb  & $0.9920,-0.1260$~\cite{Barber:2007} \\

$^{200}$Hg  & $0.9851,-0.1717$~\cite{Lurio:1965} \\

$^{24}$Mg  & $0.9970,-0.0775$~\cite{Taichenachev:2006} \\

$^{40}$Ca  & $0.9997,-0.0209$~\cite{Beverini:1998} \\

$^{112}$Cd & $0.9989,-0.0449$~\cite{Lurio:1965} \\

\hline
\hline
\end{tabular}
\caption{Coefficient mixing parameters $\alpha,\beta$ in the Russell-Saunders coupling configuration in order to evaluate the first and 2nd order Zeeman-shifts for selected atomic species.}
\label{table-IV}
\end{table}

\section*{A4. Zeeman interaction of a $^{3}$P$_{J}$ atomic state in intermediate coupling}

\subsection*{A4.1 States and matrix elements}

\indent The $|^{1,3}P_j,m_{j}\rangle$ two-electron states in intermediate coupling are written as expansions of the pure Russel-Saunders coupling states (indicated by a superscript $^{0}$) as~\cite{Lurio:1962,Boyd:2007}
\begin{equation}
\begin{split}
|^3P_2,m_{j}\rangle&=|^3P^{0}_2,m_{j}\rangle \\
|^3P_1,m_{j}\rangle&=\alpha|^3P^{0}_1,m_{j}\rangle+\beta|^1P^{0}_1,m_{j}\rangle, \\
|^1P_1,m_{j}\rangle&=-\beta|^3P^{0}_1,m_{j}\rangle+\alpha|^1P^{0}_1,m_{j}\rangle, \\
|^3P_0,0\rangle&=|^3P^{0}_0,0\rangle
\end{split}
\label{intermediatebasis}
\end{equation}
where the $(\alpha,\beta)$ parameters are reported in the Table~\ref{table-IV}.\\
\indent Because The M1 transition satisfies the $\Delta S=0$, the only magnetic contributions different from zero are within the same Russel-Saunders multiplet.
The Wigner-Eckart theorem allows us to write for the $q$ component of the magnetic dipole moment  within the Russel-Saunders basis
\begin{eqnarray}
\begin{split}
\langle n,^3P_J^0,m_{j}|d_{M1}^q|n,^{3}P_{J'}^0,m_{j}'\rangle=\langle M1\rangle\langle J,m_{j}|J',m_{j}';1,q\rangle, \nonumber
\end{split}
\end{eqnarray}
with  $\langle M1\rangle$ the reduced dipole moment and the last term representing the Clebsh-Gordan coefficient all equal 1 or the $J=0 \to J=1$ transitions of our interest. Following the treatment of a textbook as~\cite{Curtis:2001}, both the $\langle M1\rangle$ reduced dipole moment and each of the matrix element are equal to $\sqrt{2/3}$, as in~\cite{Hobson:2016,Taichenachev:2006,Boyd:2007}\\
The $\alpha,\beta$ corrections for the intermediate coupling states of Eqs.~\eqref{intermediatebasis} lead to
\begin{equation}
\begin{split}
\langle^3P_0,0|d_M|^3P_1,0\rangle&=\alpha\sqrt{\frac{2}{3}},\\
\langle^3P_0,0|d_M|^1P_1,0\rangle&=-\beta\sqrt{\frac{2}{3}}.
\end{split}
\end{equation}

\subsection*{A4.2 First-order Zeeman effect}

\indent The first-order Zeeman correction $\Delta\textup{E}^{(1)}_{Z}(^{3}P_{j})$ for a boson with a zero nuclear spin (I=0) can be evaluated in intermediate coupling by the following expression~\cite{Lurio:1962}:
\begin{equation}
\begin{split}
\Delta\textup{E}^{(1)}_{Z}(^{3}P_{J})=g'_{j}(^{3}P_{j})m_{j}\mu_{B}B
\end{split}
\label{first-Zeeman-shift}
\end{equation}
where g'$_{j}(^{3}P_{j})$ for each state of the multiplet, in a Russell-Saunders coupling configuration, are:
\begin{equation}
\begin{split}
g'_{j}(^{3}P_{2})&=g_{j}(^{3}P_{2})=3/2\\
g'_{j}(^{3}P_{1})&=\alpha^{2}g_{j}(^{3}P_{1})+\beta^{2}g_{j}(^{1}P_{1})\\
g'_{j}(^{3}P_{0})&=0
\end{split}
\label{g'_j}
\end{equation}
and:
\begin{equation}
\begin{split}
g_{j}(^{2S+1}L_{J})=&g_{l}\frac{J(J+1)+L(L+1)-S(S+1)}{2J(J+1)}\\
&+g_{s}\frac{J(J+1)+S(S+1)-L(L+1)}{2J(J+1)}
\end{split}
\label{g_j}
\end{equation}
with $g_{l}=1$ and $g_{s}=2$ is the electron spin value.
For $^{88}$Sr, we have found a linear Zeeman-shift of 21~GHz/T (2.1~MHz/G) to split each magnetic sublevel which is consistent with the experimental value reported in~\cite{Trautmann:2023}.

\subsection*{A4.3 Second-order Zeeman correction}

\indent For a boson, the 2nd-order Zeeman-shift $\Delta\textup{E}^{(2)}_{Z}(^{3}P_{j})$ is~\cite{Lurio:1962,Boyd:2007}:
\begin{equation}
\begin{split}
\Delta\textup{E}^{(2)}_{Z}(^{3}P_{J})=-\sum_{k}\frac{\left|\langle k||H_{z}||^{3}P_{J}\rangle\right|^{2}}{\textup{E}(k)-\textup{E}(^{3}P_{J})}
\end{split}
\label{2nd-Zeeman-shift}
\end{equation}
where $E(k)$ is the energy of the k-state and Zeeman matrix elements are explicitly~\cite{Lurio:1962,Boyd:2007}:
\begin{equation}
\begin{split}
\langle^{3}P_{0},0|H_{z}|^{3}P_{1},0\rangle&=\alpha(g_{s}-g_{l})\sqrt{\frac{2}{3}}\mu_{B}B\\
\langle^{3}P_{0},0|H_{z}|^{1}P_{1},0\rangle&=-\beta(g_{s}-g_{l})\sqrt{\frac{2}{3}}\mu_{B}B\\
\langle^{3}P_{2},m_{j}|H_{z}|^{3}P_{1},m_{j}\rangle&=\alpha(g_{s}-g_{l})\sqrt{\frac{4-m_{j}^{2}}{12}}\mu_{B}B\\
\langle^{3}P_{2},m_{j}|H_{z}|^{1}P_{1},m_{j}\rangle&=-\beta(g_{s}-g_{l})\sqrt{\frac{4-m_{j}^{2}}{12}}\mu_{B}B\\
\langle^{3}P_{2},0|H_{z}|^{3}P_{0},0\rangle&=0
\end{split}
\label{Zeeman-matrix-element}
\end{equation}
Again, $\alpha,\beta$ are listed in Table~\ref{table-II}. As an example, we compare our 2nd order Zeeman-shift for $\Delta m=0,\pm1$ components of the Yb magnetic quadrupole transition, to be respectively 1.2~MHz/T$^{2}$ (12~mHz/G$^{2}$) and 0.92~MHz/T$^{2}$ (9.2~mHz/G$^{2}$), consistent with theoretical values reported in~\cite{Flaumbaum:2018}.
For Zeeman sublevels $m_{j}=\pm2$, the quadratic Zeeman-shift components $\langle^{3}P_{2},m_{j}|H_{z}|^{3}P_{1},m_{j}\rangle$ and $\langle^{3}P_{2},m_{j}|H_{z}|^{1}P_{1},m_{j}\rangle$ are zero. The residual 2nd order Zeeman-shift is due to the M1 decay channels within states of different configurations. These matrix elements are small due to the orthogonality of the wave functions and suppressed by large-energy denominators~\cite{Flaumbaum:2018}. The only available value for $m_{j}=\pm2$ components of the Yb magnetic quadrupole transition was theoretically estimated by ref~\cite{Flaumbaum:2018} and is reported in Table~I. of the main text.

\section*{A5. Scalar, vector and tensor polarizabilities of $^{1}$S$_{0}$ and $^{3}$P$_{J}$ atomic states}

\indent In this subsection, we estimate the electric dipole polarizabilities for atomic states defining the E1M1 ($^{1}$S$_{0}\rightarrow^{3}$P$_{0}$) clock transition and the M2 magnetic quadrupole ($^{1}$S$_{0}\rightarrow^{3}$P$_{2}$) transition.
We use the most recent work by~\cite{Trautmann:2022-phd,Trautmann:2023} where the dynamic polarizability $\alpha_{i}$ of any atomic state $|i\rangle$ can be decomposed into a scalar polarizability $\alpha^{s}_{i}$, a vector polarizability $\alpha^{v}_{i}$ and a tensor polarizability $\alpha^{t}_{i}$ as~\cite{Trautmann:2022-phd,Trautmann:2023}:
\begin{equation}
\begin{split}
\alpha_{i}=&\alpha^{s}_{i}+\alpha^{v}_{i}\sin(2\gamma)\frac{m_{J_{i}}}{2J_{i}}\\
&+\alpha^{t}_{i}\frac{3\cos^{2}(\beta)-1}{2}\frac{3m^{2}_{J_{i}}-J_{i}(J_{i}+1)}{J_{i}(2J_{i}-1)}
\end{split}
\label{polarizability}
\end{equation}
where $\gamma$ is the ellipticity angle of the polarization and $\cos(\beta)$ is the projection of the polarization vector onto the quantization axis.
The scalar contribution of a state $|i\rangle$ with angular momentum $J_{i}$ is expressed as~\cite{Trautmann:2022-phd,Trautmann:2023}:
\begin{equation}
\begin{split}
\alpha^{s}_{i}=\frac{1}{3(2J_{i}+1)}\sum_{k}\frac{2}{\hbar}\frac{|\langle k|D|i\rangle|^{2}\omega_{ki}}{\omega^{2}_{ki}-\omega^{2}}+\alpha^{c}_{i}
\end{split}
\label{scalar-polarizability}
\end{equation}
The summation is over the dipole-allowed transitions to states $|k\rangle$ using the corresponding dipole matrix element $\langle k|D|i\rangle$.
The vector contribution is given by~\cite{Trautmann:2022-phd,Trautmann:2023}:
\begin{equation}
\begin{split}
\alpha^{v}_{i}=&-\sqrt{\frac{6J_{i}}{(J_{i}+1)(2J_{i}+1)}}\sum_{k}(-1)^{J_{i}+J_{k}}\\
&\times\left\{\begin{array}{ccc}
                                                                  1 & 1 & 1 \\
                                                                  J_{i} & J_{k} & J_{i}
                                                                \end{array}
\right\}\frac{|\langle k|D|i\rangle|^{2}\omega_{ki}}{\hbar}\left(\frac{1}{\omega_{ki}-\omega}-\frac{1}{\omega_{ki}+\omega}\right)
\end{split}
\label{vector-polarizability}
\end{equation}
and the tensor contribution is~\cite{Trautmann:2022-phd,Trautmann:2023}:
\begin{equation}
\begin{split}
\alpha^{t}_{i}=&-\sqrt{\frac{10J_{i}(2J_{i}-1)}{3(J_{i}+1)(2J_{i}+1)(2J_{i}+3)}}\sum_{k}(-1)^{J_{i}+J_{k}+1}\\
&\times\left\{\begin{array}{ccc}
                                                                  1 & 2 & 1 \\
                                                                  J_{i} & J_{k} & J_{i}
                                                                \end{array}
\right\}\frac{2}{\hbar}\frac{|\langle k|D|i\rangle|^{2}\omega_{ki}}{\omega^{2}_{ki}-\omega^{2}}
\end{split}
\label{tensor-polarizability}
\end{equation}
All required information to compute polarizabilities of $^{1}$S$_{0}$ and $^{3}$P$_{J}$ atomic states for $^{88}$Sr can be extracted from~\cite{Trautmann:2022-phd,Trautmann:2023}.
We are now able to evaluate the differential polarizability between $^{1}$S$_{0}$ and $^{3}$P$_{0}$ states for the E1M1 clock transition and the differential polarizability between $^{1}$S$_{0}$ and $^{3}$P$_{2}$ states for the M2 clock transition following the final expression~\cite{Trautmann:2022-phd,Trautmann:2023}:
\begin{equation}
\begin{split}
\Delta\alpha=&\left(\alpha^{s}_{e}+\alpha^{v}_{e}\sin(2\gamma)\frac{m_{J}}{2J}\right.\\
&\left.+\alpha^{t}_{e}\frac{3\cos^{2}(\beta)-1}{2}\frac{3m^{2}_{J}-J(J+1)}{J(2J-1)}\right)-\alpha^{s}_{g}
\end{split}
\label{differential-polarizability}
\end{equation}
The transversal and longitudinal clock light-shift are related to the differential polarizability by the expression:
\begin{equation}
\begin{split}
\Delta^{LS}_{\perp,z}(\textbf{r})\equiv\Delta^{E1}_{\perp,z}(\textbf{r})=-\frac{1}{4}\Delta\alpha\widetilde{E}^{2}_{\perp,z}(\textbf{r})
\end{split}
\label{clock-light-shift}
\end{equation}
where we have used $\widetilde{E}^{2}_{\perp,z}(\textbf{r})=2I_{\perp,z}(\textbf{r})/(c\epsilon_{0})$ and the intensity at the center of the beam given by $I_{0}=P/\pi\textup{w}^{2}_{\lambda}$ at the $\lambda$ wavelength.
The differential light-shift of the $^{1}$S$_{0}\rightarrow^{3}$P$_{0}$ clock transition is entirely scalar while vector and tensor polarizabilities give important contribution to the $^{1}$S$_{0}\rightarrow^{3}$P$_{2}$ clock transition depending on the polarization angle of the light with respect to the quantization axis.
\begin{figure*}[t!!]
\center
\resizebox{8.5cm}{!}{\includegraphics[angle=0]{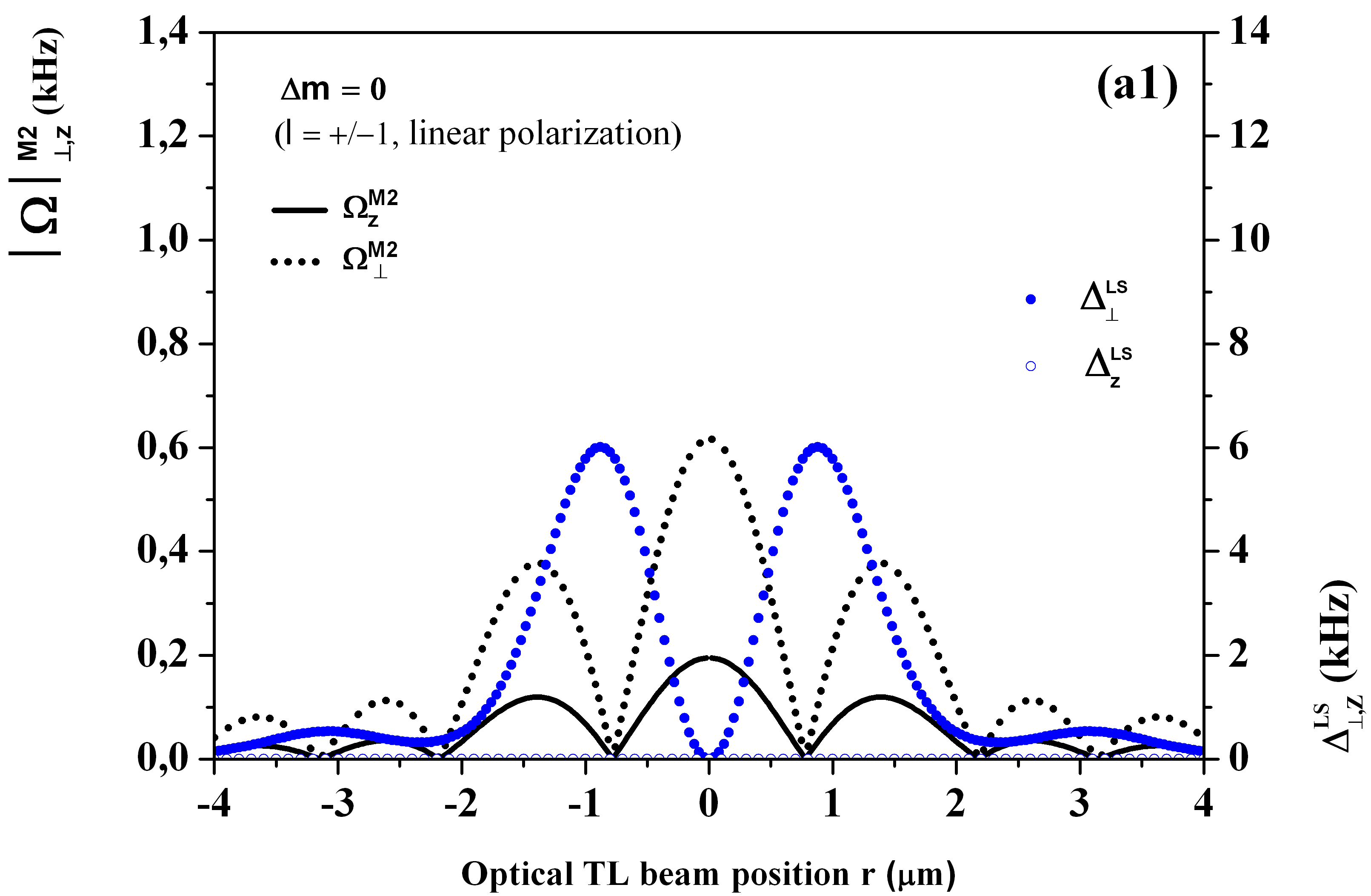}}\hspace{0.5cm}\resizebox{8.5cm}{!}{\includegraphics[angle=0]{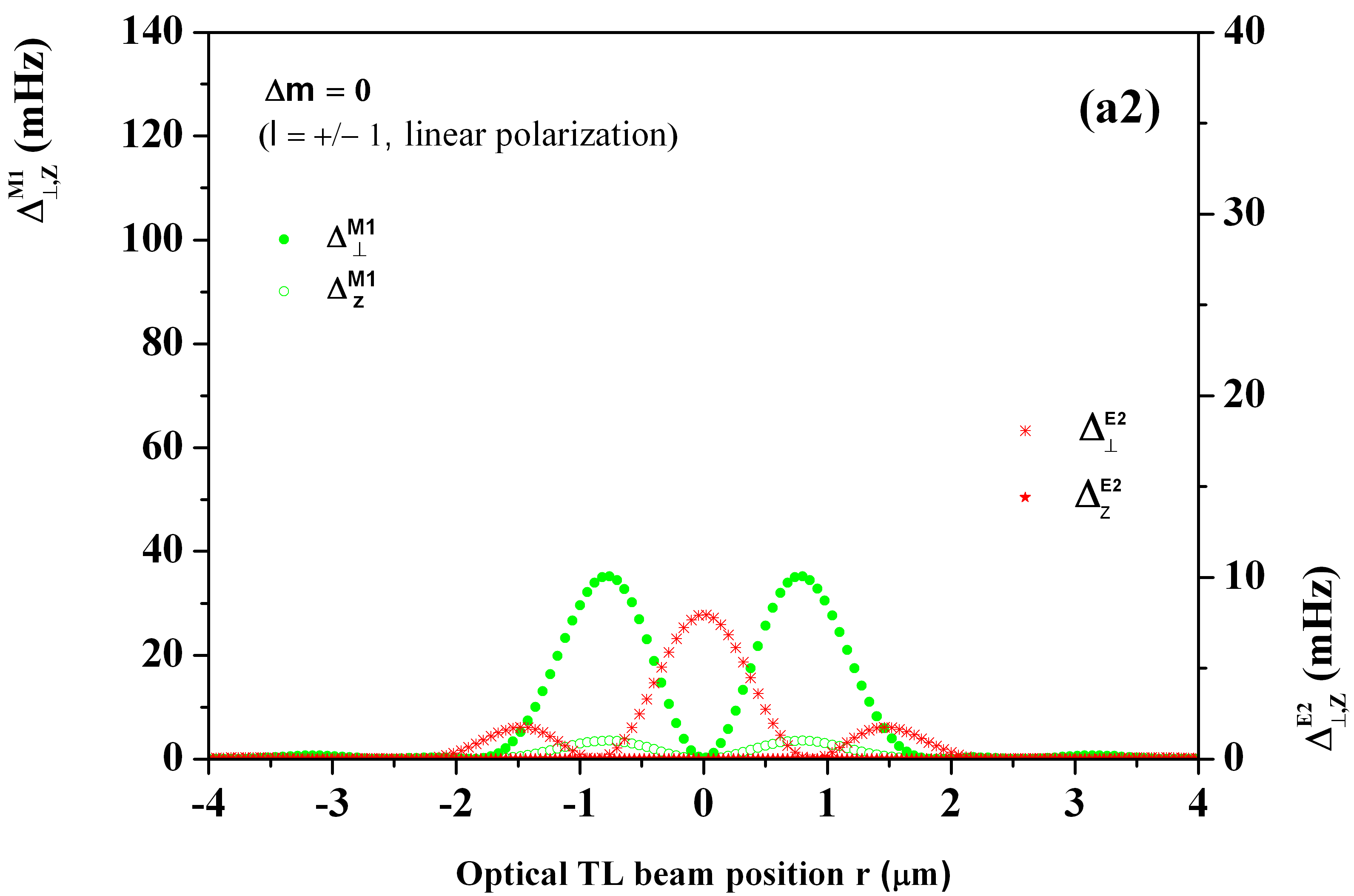}}
\resizebox{8.5cm}{!}{\includegraphics[angle=0]{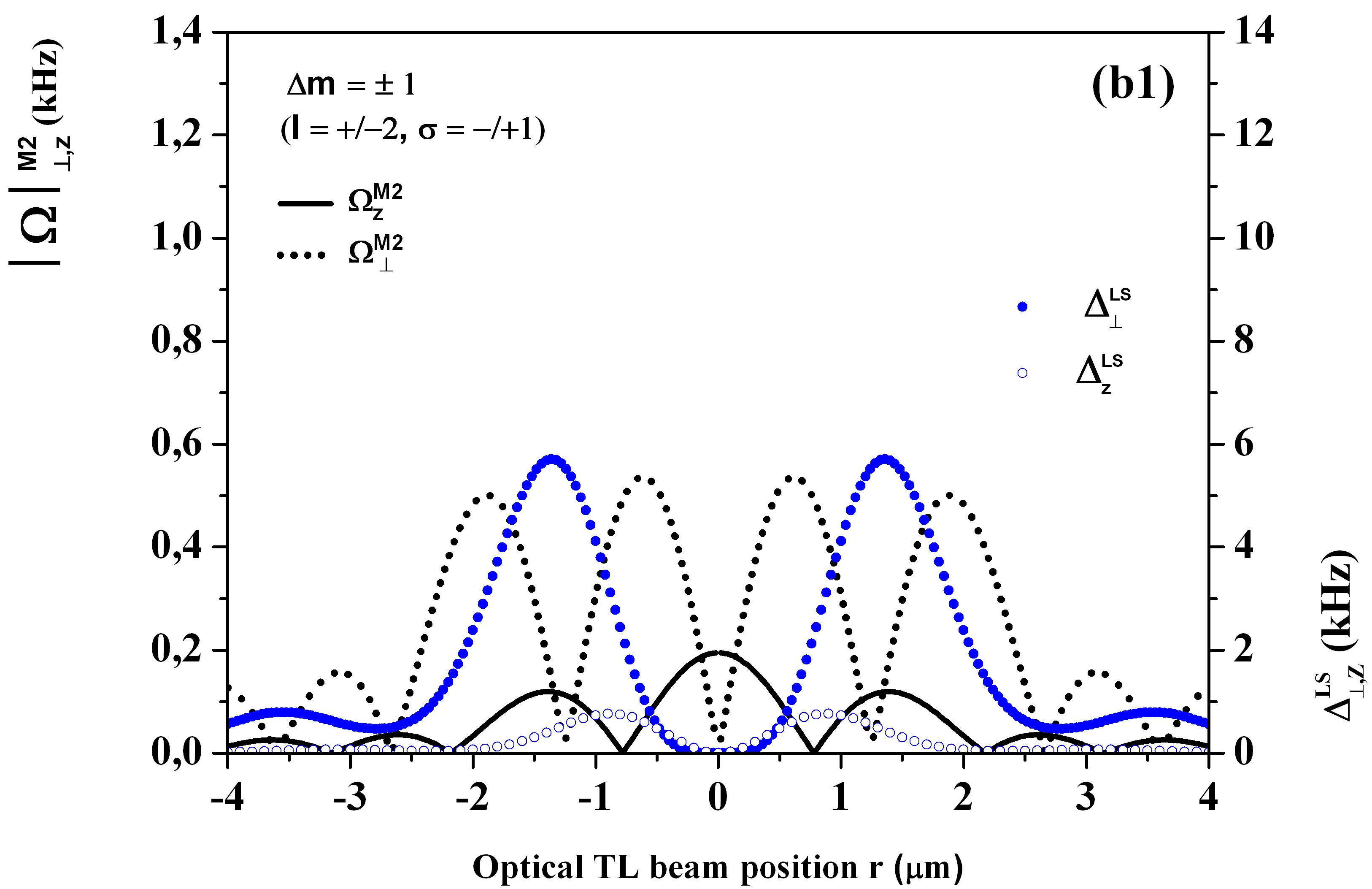}}\hspace{0.5cm}\resizebox{8.5cm}{!}{\includegraphics[angle=0]{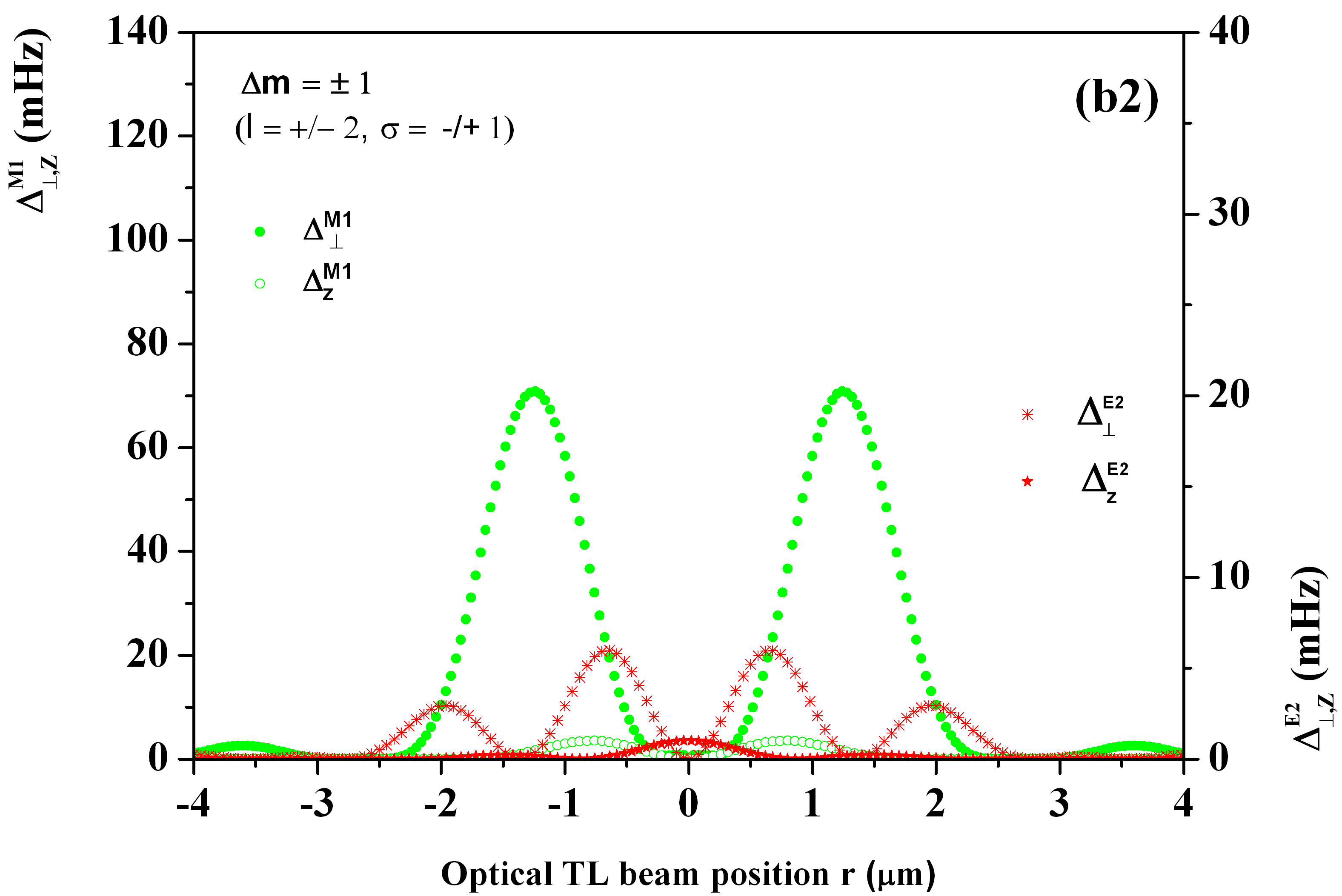}}
\resizebox{8.5cm}{!}{\includegraphics[angle=0]{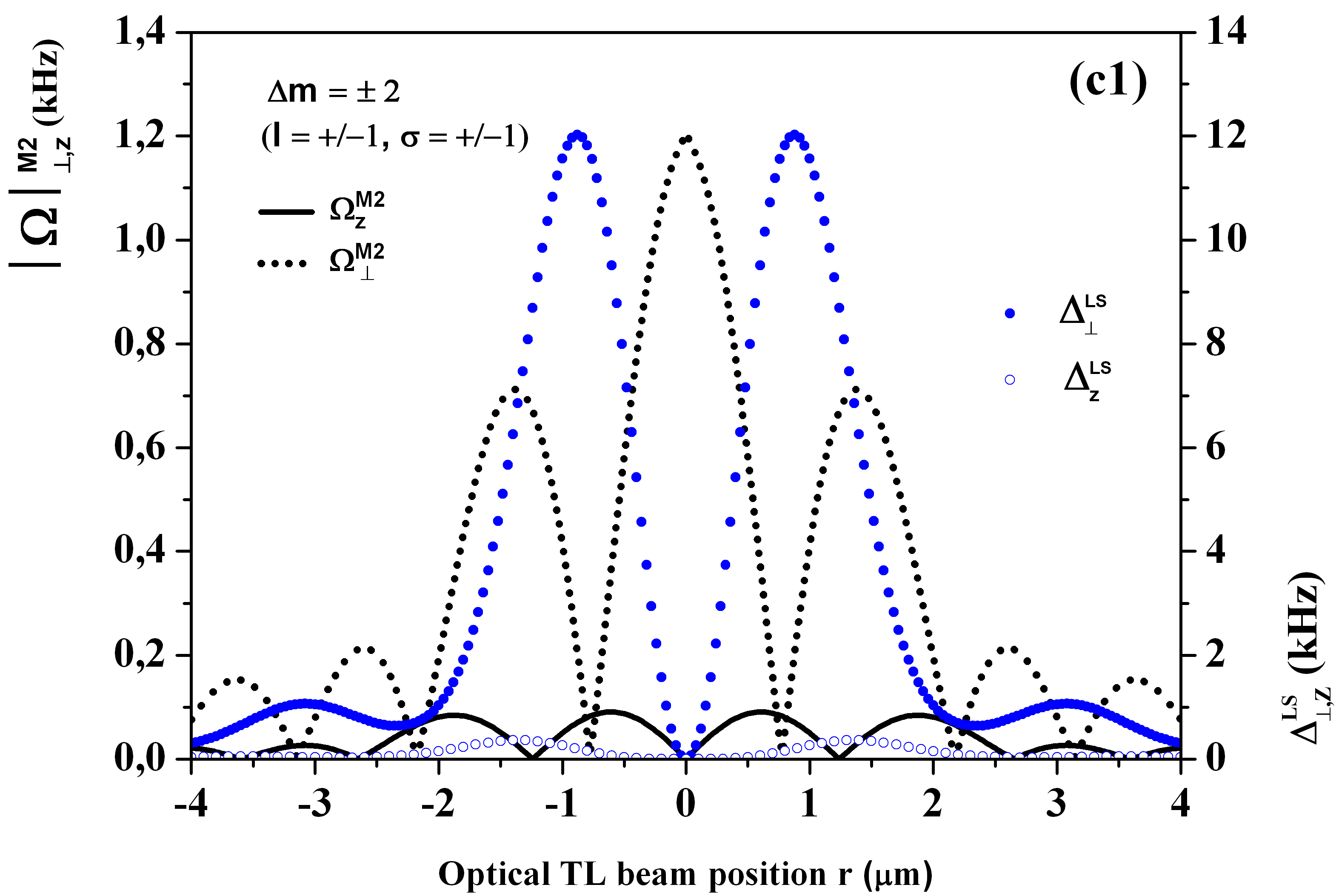}}\hspace{0.5cm}\resizebox{8.5cm}{!}{\includegraphics[angle=0]{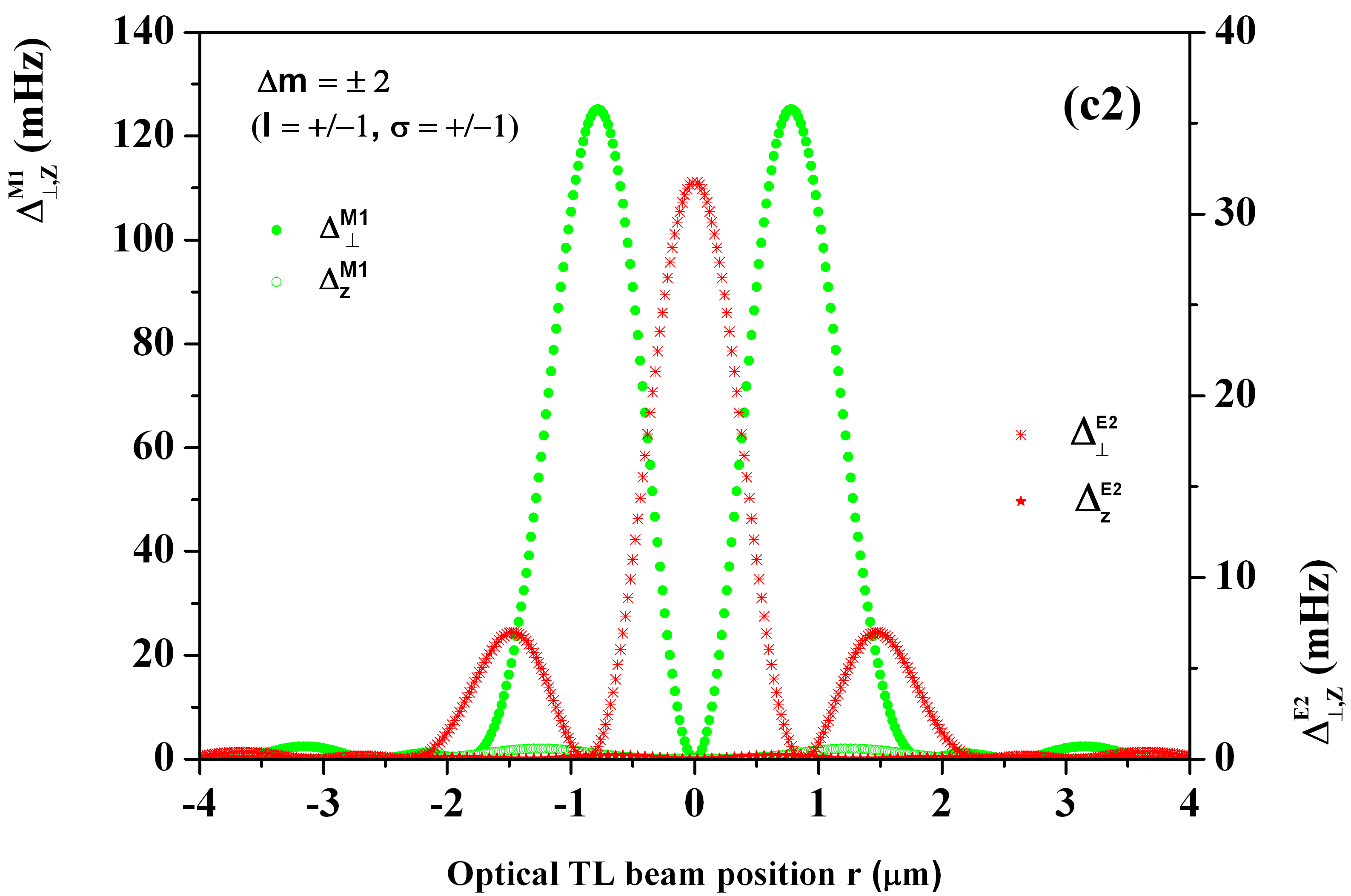}}
\caption{(color online). Excitation rates $\Omega^{M2}_{\perp,z}(\textbf{r})$ (in kHz vertical left-axis) and light-shifts $\Delta^{LS}_{\perp,z}\equiv\Delta^{E1}_{\perp,z}$ (in kHz vertical right-axis) of magnetic quadrupole Zeeman components in $^{88}$Sr. (a1) Zeeman insensitive clock component $\Delta m=0$ by a linear polarized TL beam along the Ox axis. (b1) Zeeman sensitive clock components $\Delta m=\pm1$ by a TL beam with OAM $l=\pm2$ and SAM helicity $\sigma=\mp1$. (c1) Zeeman sensitive clock components $\Delta m=\pm2$ by a TL beam with OAM $l=\pm1$ and SAM helicity $\sigma=\pm1$. (a2)-(c2) Associated multipole shifts $\Delta^{M1,E2}_{\perp,z}(\textbf{r})$ (in mHz vertical left and right-axis) of Zeeman components induced by M1 and E2 decay channels. Laser power is fixed to $\textup{P}_{671}\sim1$~mW with a waist of w$_{671}\sim2~\mu$m.}
\label{fig:M2-excitation}
\end{figure*}

\section*{A6. TL beam excitation of the $^{1}S_{0}\rightarrow^{3}P_{2}$ magnetic quadrupole components.}

\indent Driving a particular Zeeman component of a quadrupole transition by a specific TL beam is dependent on the angle orientation between the quantization axis defined by a weak external static magnetic field and the TL beam propagation axis. A TL beam can be seen as a coherent superposition of multiple plane waves with transverse and longitudinal components related to spherical harmonics where atomic selection rules are modified by spatially structured lights~\cite{Schmiegelow:2012,Schulz:2019,Schulz:2020}. For simplicity, we have ignored here this additional treatment.

We report in Fig.~\ref{fig:M2-excitation}(a1)-(c1) excitation rates $\Omega^{M2}_{\perp,z}(\textbf{r})$ of Zeeman $\Delta m=0,\pm1,\pm2$ components driven by a single TL beam versus the spatial beam position.
The magnetic quadrupole excitation rate by a TL beam is given by Eq.~\ref{M2-quadrupole-rate} where the magnetic quadrupole tensor component has been replaced by a reduced matrix element $\langle M2\rangle$ reported in Table~\ref{table-I} (Table~\ref{table-III}).

\indent $\bullet$ The magnetic-insensitive Zeeman $\Delta\textup{m}=0$ component shown in Fig.~\ref{fig:M2-excitation}(a1) can be driven by a linear polarized TL vortex beam along the Ox axis. The TL beam is decomposed into a superposition of SAM $\sigma=\pm1$ helicities coupled to OAM with l=1 by a Laguerre-Gauss (LG) or a Bessel mode where electric fields are vanishing in the beam axis center as expected~\cite{Schmiegelow:2012}. We have reported the light-shift in Fig.~\ref{fig:M2-excitation}(a1) (vertical right-axis).\\
\indent $\bullet$ The magnetic-sensitive Zeeman $\Delta\textup{m}=\pm1$ components shown in Fig.~\ref{fig:M2-excitation}(b1) can be driven by a circular polarized TL vortex beam with OAM $l=\pm2$ and SAM helicity $\sigma=\mp1$. The corresponding light-shift is reported in the same figure (vertical right-axis).\\
\indent $\bullet$ The magnetic-sensitive Zeeman $\Delta\textup{m}=\pm2$ components shown in Fig.~\ref{fig:M2-excitation}(c1) can be driven by a circular polarized TL vortex beam with OAM $l=\pm1$ and SAM helicity $\sigma=\pm1$. The corresponding light-shift is also reported in the same figure (vertical right-axis).

\indent The flexibility in addressing one particular Zeeman component of a magnetic quadrupole transition by a TL excitation is demonstrated with Fig.~\ref{fig:M2-excitation}(a1-c1). The corresponding light-shift $\Delta^{LS}_{\perp,z}(\textbf{r})\equiv\Delta^{E1}_{\perp,z}(\textbf{r})$ from off-resonant electric dipole transitions can be efficiently reduced to a vanishing level of correction by composite TL pulse protocols associated to composite pulse protocols and auto-balanced Ramsey spectroscopy~\cite{Zanon-Willette:2018,Sanner:2018,Yudin:2018}.

\section*{A7. Multipole shifts of the $^{1}S_{0}\rightarrow^{3}P_{2}$ clock transition.}

\indent We are finally looking at residual multipole shifts due to decay channels shown in Fig.~\ref{fig:decay-channels} that affect the quadrupole clock transition probed by a TL beam. We estimate reduced $\langle E2\rangle$ and $\langle M1\rangle$ coupling elements using Einstein's coefficients related to decay channels $^{3}P_{2}\rightarrow^{3}P_{0}$ and $^{3}P_{2}\rightarrow^{3}P_{1}$ reported in Table~\ref{table-III}.
By applying a second-order perturbation theory, we evaluate the magnetic dipole shift (ac Zeeman shift) $\Delta^{M1}_{\perp,z}(\textbf{r})$ based on the following expression~\cite{Beloy:2023}:
\begin{equation}
\begin{split}
\Delta^{M1}_{\perp,z}(\textbf{r})=-\frac{1}{4}\frac{|\langle^{3}P_{1}|H_{M1}|^{3}P_{2}\rangle|^{2}_{\perp,z}}{\textup{E}(^{3}P_{1})-\textup{E}(^{3}P_{2})}
\end{split}
\label{M1-light-shift}
\end{equation}
where the coupling element related to a magnetic TL excitation is given by:
\begin{equation}
\begin{split}
\langle^{3}P_{1}|H_{M1}|^{3}P_{2}\rangle_{\perp,z}=-\langle M1\rangle\cdot\int_{0}^{1}u\widetilde{B}_{\perp,z}(u\textbf{r})du
\end{split}
\label{M1-rate}
\end{equation}
The residual ac electric quadrupole shift is given by a similar expression as:
\begin{equation}
\begin{split}
\Delta^{E2}_{\perp,z}(\textbf{r})=-\frac{1}{4}\sum_{k}\frac{|\langle k|H_{E2}|^{3}P_{2}\rangle|^{2}_{\perp,z}}{\textup{E}(k)-\textup{E}(^{3}P_{2})}
\end{split}
\label{E2-light-shift}
\end{equation}
where $|k\rangle\equiv|^{3}P_{0}\rangle,|^{3}P_{1}\rangle$.
The electric quadrupole excitation rate is approximated by
\begin{equation}
\begin{split}
\langle k|H_{E2}|^{3}P_{2}\rangle_{\perp,z}=-\frac{1}{6}\langle E2\rangle\cdot\nabla_{\textbf{r}}\int_{0}^{1}\widetilde{E}_{\perp,z}(u\textbf{r})du
\end{split}
\label{E2-rate}
\end{equation}
with the electric quadrupole tensor component replaced by a reduced matrix element $\langle E2\rangle$ reported in Table~\ref{table-III}.
We have reported in Fig.~\ref{fig:M2-excitation}(a2)-(c2) the resulting clock frequency corrections from off-resonant M1 and E2 light excitations induced through decay channels.
Again, the frequency correction induced by these multipole shifts are below 150~mHz and easily reduced to a manageable level of correction by hyper-Ramsey spectroscopy~\cite{Zanon-Willette:2018}.


\begin{thebibliography}{0}

\bibitem{Ye:2008} J. Ye, H.J. Kimble and H. Katori, \textit{Quantum State Engineering and Precision Metrology Using State-Insensitive Light Traps}, \textcolor{blue}{Science \textbf{320}, 1734 (2008)}.
\bibitem{Derevianko:2011} A. Derevianko and H. Katori, \textit{Colloquium: Physics of optical lattice clocks}, \textcolor{blue}{Rev. Mod. Phys. \textbf{83}, 331 (2011)}.
\bibitem{Ludlow:2015} A.D. Ludlow, M.M. Boyd, J. Ye, E. Peik and P.O. Schmidt, \textit{Optical atomic clocks}, \textcolor{blue}{Rev. Mod. Phys. \textbf{87}, 637 (2015)}.
\bibitem{Barredo:2016} D. Barredo, S. de Léséleuc, V. Lienhard, T. Lahaye and A. Browaeys, \textit{An atom-by-atom assembler of defect-free arbitrary two-dimensional atomic arrays}, \textcolor{blue}{Science \textbf{354}, 1021 (2016)}.
\bibitem{Flaumbaum:2018} V.A. Dzuba V.V. Flambaum and S. Schiller, \textit{Testing physics beyond the standard model through additional clock transitions in neutral ytterbium}, \textcolor{blue}{Phys. Rev. A \textbf{98}, 022501 (2018)}.
\bibitem{Safronova:2018} M.S. Safronova, S.G. Porsev, C. Sanner and J. Ye, \textit{Two Clock Transitions in Neutral Yb for the Highest Sensitivity to Variations of the Fine-Structure Constant}, \textcolor{blue}{Phys. Rev. Lett. \textbf{120}, 173001 (2018)}.
\bibitem{Tang:2023} Z.-M. Tang, Y.-m. Yu, B.K. Sahoo, Ch.-Zh. Dong, Y. Yang, Y. Zou, \textit{Simultaneous magic trapping conditions for three additional clock transitions in Yb to search for variation of the fine-structure constant}, \textcolor{blue}{Phys. Rev. A \textbf{107}, 053111 (2023)}.
\bibitem{Ishiyama:2023} T. Ishiyama, K. Ono, T. Takano, A. Sunaga and Y. Takahashi, \textit{Observation of an Inner-Shell Orbital Clock Transition in Neutral Ytterbium Atoms}, \textcolor{blue}{Phys. Rev. Lett. (2023)}.
\bibitem{Boyd:2007} M.M. Boyd, T. Zelevinsky, A.D. Ludlow, S. Blatt, T. Zanon-Willette, S.M. Foreman and J. Ye, \textit{Nuclear spin effects in optical lattice clocks}, \textcolor{blue}{Phys. Rev. A \textbf{76}, 022510 (2007)}.
\bibitem{Taichenachev:2006} A.V. Taichenachev, V.I. Yudin, C.W. Oates, C.W. Hoyt, Z.W. Barber and L. Hollberg, \textit{Magnetic Field-Induced Spectroscopy of Forbidden Optical Transitions
with Application to Lattice-Based Optical Atomic Clocks}, \textcolor{blue}{Phys. Rev. Lett. \textbf{96}, 083001 (2006)}.
\bibitem{Barber:2006} Z.W. Barber, C.W. Hoyt, C.W. Oates, L. Hollberg, A.V. Taichenachev and V.I. Yudin, \textit{Direct Excitation of the Forbidden
Clock Transition in Neutral $^{174}$Yb Atoms Confined to an Optical Lattice}, \textcolor{blue}{Phys. Rev. Lett. \textbf{96}, 083002 (2006)}.
\bibitem{Baillard:2007} X. Baillard, M. Fouché, R. Le Targat, P.G. Westergaard, A. Lecallier, Y. Le Coq, G.D. Rovera, S. Bize and P. Lemonde, \textit{Accuracy evaluation of an optical lattice clock with bosonic atoms}, \textcolor{blue}{Opt. Lett. \textbf{32}, 1812 (2007)}.
\bibitem{Kulosa:2015} A.P. Kulosa, D. Fim, K.H. Zipfel, S. Rühmann, S. Sauer, N. Jha, K. Gibble, W. Ertmer, E.M. Rasel, M.S. Safronova, U.I. Safronova and S.G. Porsev, \textit{Towards a Mg Lattice Clock:
Observation of the $^{1}$S$_{0}$-$^{3}$P$_{0}$ Transition and Determination of the Magic Wavelength}, \textcolor{blue}{Phys. Rev. Lett. \textbf{115}, 240801 (2015)}.
\bibitem{Daley:2008} A.J. Daley, M.M. Boyd, J.Ye and P. Zoller, \textit{Quantum computing with alkaline-earth-metal atoms}, \textcolor{blue}{Phys. Rev. Lett. \textbf{101}, 170504 (2008)}.
\bibitem{Kato:2012} S. Kato, K. Shibata, R. Yamamoto, Y. Yoshikawa and Y. Takahashi, \textit{Optical magnetic resonance imaging with an ultra-narrow optical transition}, \textcolor{blue}{Appl. Phys. B \textbf{108}, 31 (2012)}.
\bibitem{Onishchenko:2019} O. Onishchenko, S. Pyatchenkov, A. Urech, Ch.-Ch. Chen, S. Bennetts, G.A. Siviloglou and F. Schreck, \textit{Frequency of the ultranarrow $^{1}$S$_{0}$-$^{3}$P$_{2}$ transition in $^{87}$Sr}, \textcolor{blue}{Phys. Rev. A \textbf{99}, 052503 (2019)}.
\bibitem{Trautmann:2023} J. Trautmann, D. Yankelev, V. Klüsener, A. J. Park, I. Bloch and S. Blatt, \textit{The $^{1}$S$_{0}$-$^{3}$P$_{2}$ magnetic quadrupole transition in neutral strontium}, \textcolor{blue}{Phys. Rev. Research. \textbf{5}, 013219 (2023)}.
\bibitem{Santra:2005} R. Santra, E. Arimondo, T. Ido, C. Greene and J. Ye, \textit{High-accuracy optical clock via three-level coherence in neutral bosonic $^{88}$Sr}, \textcolor{blue}{Phys. Rev. Lett. \textbf{94}, 173002 (2005)}.
\bibitem{Beloy:2021} K. Beloy, \textit{Prospects of a Pb$^{2+}$ Ion Clock}, \textcolor{blue}{Phys. Rev. Lett. \textbf{127}, 013201 (2021)}.
\bibitem{Alden:2014} E.A. Alden, K.R. Moore and A.E. Leanhardt, \textit{Two-photon E1-M1 optical clock}, \textcolor{blue}{Phys. Rev. A \textbf{90}, 012523 (2014)}.
\bibitem{Schulz:2020} S.A.-L. Schulz, A.A. Peshkov, R.A. Müller, R. Lange, N. Huntemann, Chr. Tamm, E. Peik and A. Surzhykov, \textit{Generalized excitation of atomic multipole transitions by twisted light modes}, \textcolor{blue}{Phys. Rev. A \textbf{102}, 012812 (2020)}.
\bibitem{Lange:2022} R. Lange, N. Huntemann , A.A. Peshkov , A. Surzhykov and E. Peik, \textit{Excitation of an Electric Octupole Transition by Twisted Light}, \textcolor{blue}{Phys. Rev. Lett. \textbf{129}, 253901 (2022)}.
\bibitem{Verde:2023} M. Verde, C.T. Schmiegelow, U. Poschinger and F. Schmidt-Kaler, \textit{Trapped atoms in spatially-structured vector light fields}, \textcolor{blue}{\textbf{arXiv:2306.17571} (2023)}.
\bibitem{Monteiro:2009} P.B. Monteiro, P.A. Maia Neto, and H. Moysés Nussenzveig \textit{Angular momentum of focused beams: Beyond the paraxial approximation}, \textcolor{blue}{Phys. Rev. A \textbf{79},  033830 (2009)}.
\bibitem{Klimov:2012} V.V. Klimov, D. Bloch, M. Ducloy and J.R. Rios Leite, \textit{Mapping of focused Laguerre-Gauss beams: The interplay between spin and orbital angular momentum and its dependence on detector characteristics}, \textcolor{blue}{Phys. Rev. A \textbf{85}, 053834 (2012)}.
\bibitem{Quinteiro:2017} G.F. Quinteiro, D.E. Reiter and T. Kuhn, \textit{Formulation of the twisted-light-matter interaction at the phase singularity:
Beams with strong magnetic fields}, \textcolor{blue}{Phys. Rev. A \textbf{95}, 012106 (2017)}.
\bibitem{Quinteiro:2023} G.F. Quinteiro Rosen, \textit{On the importance of the longitudinal component of paraxial optical vortices in the interaction with atoms}, \textcolor{blue}{J. Opt. Soc. Am. B \textbf{40}, C73 (2023)}.
\bibitem{Schmiegelow:2012} C.T. Schmiegelow and F. Schmidt-Kaler, \textit{Light with orbital angular momentum interacting with trapped ions}, \textcolor{blue}{Eur. J. phys. D \textbf{66}, 157 (2012)}.
\bibitem{Schmiegelow:2016} C.T. Schmiegelow, J. Schulz, H. Kaufmann, T. Ruster, U.G. Poschinger and F. Schmidt-Kaler, \textit{Transfer of optical orbital angular momentum to a bound electron}, \textcolor{blue}{Nat Commun \textbf{7}, 12998 (2016)}.
\bibitem{Quinteiro:2017-bis} G.F. Quinteiro, F. Schmidt-Kaler and C.T. Schmiegelow, \textit{Twisted-light-ion interaction: The role of longitudinal fields}, \textcolor{blue}{Phys. Rev. Lett. \textbf{119}, 253203 (2017)}.
\bibitem{Shen:2019} Y. Shen, X. Wang, Z. Xie, C. Min, X. Fu, Q. Liu, M. Gong and X. Yuan , \textit{Optical vortices 30 years on: OAM manipulation from topological charge to multiple singularities}, \textcolor{blue}{ Light Sci Appl \textbf{8}, 90 (2019)}.
\bibitem{Babiker:2019} M. Babiker, D.L. Andrews and V.E. Lembessis, \textit{Atoms in complex twisted light}, \textcolor{blue}{J. Opt. \textbf{21}, 013001 (2019)}.

\bibitem{Zanon-Willette:2018} T. Zanon-Willette, R. Lefevre, R. Metzdorff, N. Sillitoe, S. Almonacil, M. Minissale, E. de Clercq,
A.V. Taichenachev, V.I. Yudin and E. Arimondo, \textit{Composite laser-pulses spectroscopy for high-accuracy optical clocks: a review of recent progress and perspectives}, \textcolor{blue}{Rep. Prog. Phys. \textbf{81}, 094401 (2018)}.

\bibitem{Yudin:2011}  V.I. Yudin,A.V. Taichenachev, M.V. Okhapkin, S.N. Bagayev, Chr. Tamm, N. Huntemann, T.E. Mehlstäubler and F. Riehle, \textit{Atomic Clocks with Suppressed Blackbody Radiation Shift}, \textcolor{blue}{Phys. Rev. Lett. \textbf{107}, 030801 (2011)}.
\bibitem{Yudin:2021}  V.I. Yudin,A.V. Taichenachev, M.Yu. Basalaev, O.N. Prudnikov, H.A. Fürst, T.E. Mehlstäubler and S.N. Bagayev, \textit{Combined atomic clock with blackbody-radiation-shift-induced instability below $10^{-19}$ under natural environment conditions}, \textcolor{blue}{New J. Phys. \textbf{23}, 023032 (2021)}.
\bibitem{Young:2020} A.W. Young W.J. Eckner, W.R. Milner, D. Kedar, M.A. Norcia, E. Oelker, N. Schine, J. Ye and A.M. Kaufman, \textit{Half-minute-scale atomic coherence and high relative stability in a tweezer clock}, \textcolor{blue}{Nature \textbf{588}, 408 (2020)}.
\bibitem{Kaufman:2021} A.M. Kaufman and K.-K. Ni, \textit{Quantum science with optical tweezer arrays of ultracold atoms and molecules}, \textcolor{blue}{Nat. Phys. \textbf{17}, 1324 (2021)}.
\bibitem{Tian:2023} W. Tian, W.J. Wee, A. Qu, B.J. Ming Lim, P.R. Datla, V.P. Wen Koh, and H. Loh, \textit{Parallel Assembly of Arbitrary Defect-Free Atom Arrays with a Multitweezer Algorithm}, \textcolor{blue}{Phys. Rev. Applied \textbf{19}, 034048 (2023)}.
\bibitem{Derevianko:2001} A. Derevianko, \textit{Feasibility of Cooling and Trapping Metastable Alkaline-Earth Atoms}, \textcolor{blue}{Phys. Rev. Lett. \textbf{87}, 023002 (2001)}.
\bibitem{Goppert-Mayer:1931} M. Göppert-Mayer, \textit{Über elementarakte mit zwei quantum sprüngen}, \textcolor{blue}{Annalen der Physik \textbf{9}, 273 (1931)}.
\bibitem{Jackson:2019} S. Jackson and A.C. Vutha, \textit{Magic polarization for cancellation of light shifts in two-photon optical clocks}, \textcolor{blue}{Phys. Rev. A \textbf{99}, 063422 (2019)}.
\bibitem{Porsev:2001} S.G. Porsev, M.G. Kozlov, Yu.G. Rakhlina and A. Derevianko, \textit{Many-body calculations of electric-dipole amplitudes for transitions between low-lying levels of Mg, Ca, and Sr}, \textcolor{blue}{Phys. Rev. A \textbf{64}, 012508 (2001)}.
\bibitem{Porsev:1999} S.G. Porsev, Yu. G. Rakhlina and M.G. Kozlov, \textit{Electric-dipole amplitudes, lifetimes, and polarizabilities of the low-lying levels of atomic ytterbium}, \textcolor{blue}{Phys. Rev. A \textbf{60}, 2781 (1999)}.
\bibitem{Petersen:2008} M. Petersen, R. Chicireanu, S.T. Dawkins, D.V. Magalhães, C. Mandache, Y. Le Coq, A. Clairon and S. Bize, \textit{Doppler-Free Spectroscopy of the $^{1}$S$_{0}$-$^{3}$P$_{0}$ Optical Clock Transition in Laser-Cooled Fermionic Isotopes of Neutral Mercury}, \textcolor{blue}{Phys. Rev. Lett. \textbf{101}, 183004  (2008)}.
\bibitem{Yamaguchi:2019} A. Yamaguchi, M.S. Safronova, K. Gibble and H. Katori, \textit{Narrow-line Cooling and Determination of the Magic Wavelength of Cd}, \textcolor{blue}{Phys. Rev. Lett. \textbf{123}, 113201  (2019)}.
\bibitem{Garstang:1967} R.H. Garstang, \textit{Magnetic quadrupole line intensities}, \textcolor{blue}{Astrophysical Journal \textbf{148}, 579 (1967)}.
\bibitem{Afanasev:2013} A. Afanasev, C.E. Carlson and A. Mukherjee, \textit{Off-axis excitation of hydrogenlike atoms by twisted photons}, \textcolor{blue}{Phys. Rev. A \textbf{88}, 033841 (2013)}.
\bibitem{Scholz-Marggraf:2014} H.M. Scholz-Marggraf, S. Fritzsche, V.G. Serbo, A. Afanasev and A. Surzhykov, \textit{Absorption of twisted light by hydrogenlike atoms}, \textcolor{blue}{Phys. Rev. A \textbf{90}, 013425 (2014)}.
\bibitem{Xu:2010} P. Xu, X. He, J. Wang and M. Zhan, \textit{Trapping a single atom in a blue detuned optical bottle beam trap}, \textcolor{blue}{Opt. Lett. \textbf{35}, 2164 (2010)}.
\bibitem{Chalony:2011} M. Chalony, A. Kastberg, B. Klappauf and D. Wilkowski, \textit{Doppler Cooling to the Quantum Limit}, \textcolor{blue}{Phys. Rev. Lett. \textbf{107}, 243002 (2011)}.
\bibitem{Sanner:2018} Ch. Sanner, N. Huntemann, R. Lange, Ch. Tamm and E. Peik, \textit{Autobalanced Ramsey Spectroscopy}, \textcolor{blue}{Phys. Rev. Lett. \textbf{120}, 053602 (2018).}
\bibitem{Yudin:2018} V.I. Yudin, A.V. Taichenachev, M. Yu. Basalaev, T. Zanon-Willette, J.W. Pollock, M. Shuker, E.A. Donley and J. Kitching, \textit{Generalized Autobalanced Ramsey Spectroscopy of Clock Transitions}, \textcolor{blue}{Phys. Rev. Applied. \textbf{120}, 054034 (2018).}
\bibitem{Bohman:2023} M.A. Bohman, S.G. Porsev, D.B. Hume, D.R. Leibrandt and M.S. Safronova, \textit{Enhancing Divalent Optical Atomic Clocks with the $^{1}S_{0}\leftrightarrow^{3}P_{2}$ Transition}, \textcolor{blue}{\textbf{arXiv:2308.02056} (2023)}.
\bibitem{Nemirovsky:2023} J. Nemirovsky, R. Weill, I. Meltzer and Y. Sagi, \textit{Atomic interferometer based on optical tweezers}, \textcolor{blue}{\textbf{arXiv:2308.07768} (2023)}.
\bibitem{Premawardhana:2023} G. Premawardhana,1, J. Kunjummen, S. Subhankar and J.M. Taylor, \textit{Investigating the feasibility of a trapped atom interferometer with movable traps}, \textcolor{blue}{\textbf{arXiv:2308.12246} (2023)}.
\bibitem{Quinteiro:2015} G.F. Quinteiro, D.E. Reiter and T. Kuhn, \textit{Formulation of the twisted-light-matter interaction at the phase singularity: The twisted-light gauge}, \textcolor{blue}{Phys. Rev. A \textbf{91}, 033808 (2015)}.
\bibitem{Koksal:2022} K. Koksal, M. Babiker, V.E. Lembessis and J. Yuan, \textit{Hopf index and the helicity of elliptically polarized twisted light}, \textcolor{blue}{J. Opt. Soc. Am. B \textbf{39}, 459 (2022)}.
\bibitem{Hobson:2016} R. Hobson, \textit{An Optical Lattice Clock with Neutral Strontium}, \textcolor{blue}{Thesis (Ph.D.) Balliol College, University of Oxford, (2016)}.
\bibitem{Mizushima:1964} M. Mizushima, \textit{$\Delta S=\pm1$ Magnetic Multipole Radiative Transitions}, \textcolor{blue}{Phys. Rev. \textbf{134}, 883 (1964)}.
\bibitem{Mizushima:1966} M. Mizushima, \textit{$\Delta S=\pm1$ Magnetic Quadrupole Radiative Transitions in Atoms and Molecules}, \textcolor{blue}{J. Phys. Soc. Jpn. \textbf{21}, 2335 (1966)}.
\bibitem{Lurio:1962} A. Lurio, M. Mandel and R. Novick, \textit{Second-order hyperfine and Zeeman corrections for an (sl) configuration}, \textcolor{blue}{Phys. Rev. \textbf{126}, 1758 (1962)}.
\bibitem{Curtis:2001} L.J. Curtis, \textit{Atomic Structure and Lifetimes: a conceptual approach}, (Cambridge University Press, Cambridge, 2001).
\bibitem{Barber:2007} Z. Barber, \textit{Ytterbium Optical Lattice Clock}, \textcolor{blue}{Thesis (Ph.D.) University of Colorado at Boulder, (2007)}.
\bibitem{Lurio:1965} A. Lurio, \textit{Configuration Interaction and the hfs of the sl Configuration}, \textcolor{blue}{Phys. Rev. \textbf{142}, 46 (1965)}.
\bibitem{Beverini:1998} N. Beverini, E. Maccioni and F. Strumia, \textit{g$_{j}$ factor of neutral calcium $^{3}$P metastable levels}, \textcolor{blue}{J. Opt. Soc. Am. B \textbf{15}, 2206 (1998)}.
\bibitem{Trautmann:2022-phd} J. Trautmann, \textit{The Magnetic Quadrupole Transition in Neutral Strontium}, \textcolor{blue}{Thesis (Ph.D.) Dissertation an der Fakultät für Physik Ludwig-Maximilians-Universität München, (2022)}.
\bibitem{Schulz:2019} S.A.-L. Schulz, S. Fritzsche, R.A. Müller and A. Surzhykov, \textit{Modification of multipole transitions by twisted light}, \textcolor{blue}{Phys. Rev. A \textbf{100}, 043416 (2019)}.
\bibitem{Santra:2004} R. Santra, K.V. Christ and C.H. Green, \textit{Properties of metastable alkaline-earth-metal atoms calculated using an accurate effective core potential}, \textcolor{blue}{Phys. Rev. A \textbf{69}, 042510 (2004)}.
\bibitem{Beloy:2023} K. Beloy, \textit{Trap-Induced ac Zeeman Shift of the Thorium-229 Nuclear Clock Frequency}, \textcolor{blue}{Phys. Rev. Lett. \textbf{130}, 103201 (2023)}.
\bibitem{Jensen:2011} B.B. Jensen, He Ming, P.G.Westergaard, K. Gunnarsson, M.H. Madsen, A. Brusch, J. Hald and J.W. Thomsen, \textit{Experimental Determination of the $^{24}$Mg I (3s3p) $^{3}$P$_{2}$ Lifetime}, \textcolor{blue}{Phys. Rev. Lett. \textbf{107}, 113001 (2011)}.
\bibitem{Lu:2022} B. Lu, X. Lu, J. Li and H. Chang, \textit{Reconciliation of Theoretical Lifetimes of the $5s5p$ $^{3}$P$_{2}$ Metastable State for $^{88}$Sr with Measurement: The Role of the Blackbody-Radiation-Induced Decay}, \textcolor{blue}{Chin. Phys. Lett. \textbf{107}, 073201 (2022)}.
\end{thebibliography}
\end{document}